\documentclass[12pt,preprint]{aastex}
\slugcomment{Accepted for publication in ApJ}
\usepackage{rotating}
\usepackage{graphics}
\usepackage{graphicx}
\usepackage{amssymb}

\shorttitle{Adiabatic Mass Loss}
\shortauthors{Ge et al.}

\begin{document}
\title{Adiabatic Mass Loss in Binary Stars - I. Computational Method}

\author{Hongwei Ge\altaffilmark{1,2,3}, Michael S. Hjellming\altaffilmark{4}, Ronald F. Webbink\altaffilmark{5}, Xuefei Chen\altaffilmark{1,2}, and Zhanwen Han\altaffilmark{1,2}}
\altaffiltext{1}{National Astronomical Observatories / Yunnan Observatory, the Chinese Academy of Sciences, Kunming, 650011, China; hongwei.ge@gmail.com}
\altaffiltext{2}{Key Laboratory for the Structure and Evolution of Celestial Objects, Chinese Academy of Sciences, Kunming 650011, China}
\altaffiltext{3}{Graduate University of Chinese Academy of Sciences, Beijing 100039, China}
\altaffiltext{4}{Cherryville, North Carolina 28021, U.S.A; mshjell@gmail.com}
\altaffiltext{5}{Department of Astronomy, University of Illinois, 1002 W. Green St. Urbana, IL 61801, U.S.A; webbink@astro.uiuc.edu}

\begin{abstract}
The asymptotic response of donor stars in interacting binary systems to very rapid mass loss is characterized by adiabatic expansion throughout their interiors. In this limit, energy generation and heat flow through the stellar interior can be neglected. We model this response by constructing model sequences, beginning with a donor star filling its Roche lobe at an arbitrary point in its evolution, holding its specific entropy and composition profiles fixed as mass is removed from the surface. The stellar interior remains in hydrostatic equilibrium. Luminosity profiles in these adiabatic models of mass-losing stars can be reconstructed from the specific entropy profiles and their gradients. These approximations are validated by comparison with time-dependent binary mass transfer calculations. We describe how adiabatic mass loss sequences can be used to quantify threshold conditions for dynamical time scale mass transfer, and to establish the range of post-common envelope binaries that are allowed energetically.

In dynamical time scale mass transfer, the adiabatic response of the donor star drives it to expand beyond its Roche lobe, leading to runaway mass transfer and the formation of a common envelope with its companion star. For donor stars with surface convection zones of any significant depth, this runaway condition is encountered early in mass transfer, if at all; but for main sequence stars with radiative envelopes, it may be encountered after a prolonged phase of thermal time scale mass transfer, a so-called delayed dynamical instability. We identify the critical binary mass ratio for the onset of dynamical time scale mass transfer as that ratio for which the adiabatic response of the donor star radius to mass loss matches that of its Roche lobe at some point during mass transfer; if the ratio of donor to accretor masses exceeds this critical value, dynamical time scale mass transfer ensues.

In common envelope evolution, the dissipation of orbital energy of the binary provides the energy to eject the common envelope; the energy budget for this process consists essentially of the initial orbital energy of the binary and the initial self-energies of the binary components. We emphasize that, because stellar core and envelope contribute mutually to each other's gravitational potential energy, proper evaluation of the total energy of a star requires integration over the entire stellar interior, not the ejected envelope alone as commonly assumed. We show that the change in total energy of the donor star, as a function of its remaining mass along an adiabatic mass-loss sequence, can be calculated either by integration over initial and final models, or by a path integral along the mass-loss sequence. That change in total energy of the donor star, combined with the requirement that both remnant donor and its companion star fit within their respective Roche lobes, then circumscribes energetically possible survivors of common envelope evolution.
\end{abstract}

\keywords{binaries: close --- stars: evolution --- stars: interiors --- stars: mass loss}

\section{Introduction}
\label{intro}

Mass transfer is the signature feature of close binary evolution. In relatively wide binaries, this mass transfer can occur via capture of some part of the stellar wind of the donor by the accretor, but in close systems, the dominant mode of mass transfer is by Roche lobe overflow (RLOF). Hydrodynamic estimates of Roche lobe overflow rates \citep{pacz72,egg06} show that mass transfer rates are typically of order
\begin{equation}
\dot{M}_1 \approx -\frac{M_1}{P_{\rm orb}} \left( \frac{\Delta R_1}{R_{\rm L}} \right)^{n + \frac{3}{2}} \ ,
\label{Mdot}
\end{equation}
where $M_1$ is the mass of the donor star, $P_{\rm orb}$ the orbital period of the binary, $R_{\rm L}$ the effective Roche lobe radius, which is the radius of a sphere of the same volume as the interior of the Roche lobe, $\Delta R_1 = R_1 - R_{\rm L}$ the donor's radius $R_1$ minus its Roche lobe radius $R_{\rm L}$, and $n$ the donor's effective polytropic index. This mass transfer rate is exquisitely sensitive to $\Delta R_1$; so long as mass transfer occurs on a time scale much longer than the dynamical time scale of the binary, $R_1 \approx R_{\rm L}$.

We can examine the stability of mass transfer by comparing the responses of donor star radius and Roche lobe radius to perturbations in the mass transfer rate. So long as $\Delta R_1 \ll R_{\rm L}$, hydrostatic equilibrium prevails through the interior of the donor star, and dynamical terms become important only in the vicinity of the inner Lagrangian point. Variations in the radius of the donor can be separated into mass- and time-dependent parts:
\begin{equation}
 {\rm d} \ln R_1 = \left( \frac{\partial \ln R_1}{\partial t} \right)_{M_1} {\rm d}t + \left( \frac{\partial \ln R_1}{\partial \ln M_1} \right)_t {\rm d} \ln M_1 \ .
\label{dlnR1}
\end{equation}
The first term on the right-hand side represents not only the evolutionary expansion (or contraction) of the donor star in the absence of mass transfer, but also the expansion or contraction due to thermal relaxation if, for example, mass transfer has driven the donor out of thermal equilibrium. The second term on the right-hand side represents the hydrostatic response of the radius to changing mass, with thermal or chemical diffusion and nuclear evolution suppressed; that is, it represents the adiabatic response of the donor star to mass loss. We will henceforth label this logarithmic derivative by the subscript `ad' instead of $t$, and define
\begin{equation}
\zeta_{\rm ad} \equiv \left( \frac{\partial \ln R_{1}}{\partial \ln M_{1}} \right)_t \ .
\label{zetaad}
\end{equation}
The Roche lobe radius of the donor depends on three quantities (assuming a circular orbit) -- its mass $M_1$, the total mass of the binary $M$, and the orbital angular momentum of the binary $J$:
\begin{equation}
R_{\rm L} = r_{\rm L}(M_1/M) \frac{J^2M}{G M_1^2(M - M_1)^2} \ ,
\label{radiuslobe1}
\end{equation}
where $r_{\rm L}(M_1/M)$ is the Roche lobe radius in units of the binary separation, and is a function only of the binary mass ratio, or as written here, of the ratio of donor to total mass. We can thus write
\begin{equation}
{\rm d} \ln R_{\rm L} = \left( \frac{\partial \ln R_{\rm L}}{\partial \ln M_1} \right)_{M,J} {\rm d} \ln M_1 + \left( \frac{\partial \ln R_{\rm L}}{\partial \ln M} \right)_{M_1,J} {\rm d} \ln M + \left( \frac{\partial \ln R_{\rm L}}{\partial \ln J} \right)_{M_1,M} {\rm d} \ln J \ .
\label{dlnRL_mmj}
\end{equation}
In conservative mass transfer, as that term is commonly used, $M$ and $J$ are conserved quantities, and therefore the latter two terms on the right-hand side of Eq.~(\ref{dlnRL_mmj}) vanish. However, more generally, a binary may suffer background mass loss through a stellar wind from the companion star, or angular momentum losses through, for example, stellar winds or gravitational wave radiation. Moreover, systemic mass and angular momentum losses may contain contributions arising from a loss of part of the mass transfer stream between binary components; we can call these consequential losses. We therefore decompose the terms in ${\rm d} \ln M$ and ${\rm d} \ln J$ into terms linearly independent of ${\rm d} \ln M_1$ and given the subscript $M_1$, and those consequential losses linearly proportional to ${\rm d} \ln M_1$ and given subscript $t$, by analogy with Eq.~(\ref{dlnR1}). We then write
\begin{equation}
{\rm d} \ln R_{\rm L} = \left( \frac{\partial \ln R_{\rm L}}{\partial t} \right)_{M_1} {\rm d}t + \left( \frac{\partial \ln R_{\rm L}}{\partial \ln M_1} \right)_t {\rm d} \ln M_1 \ ,
\label{dlnRL}
\end{equation}
where
\begin{equation}
\left( \frac{\partial \ln R_{\rm L}}{\partial t} \right)_{M_1} = \left( \frac{\partial \ln R_{\rm L}}{\partial \ln M} \right)_{M_1,J} \left( \frac{\partial \ln M}{\partial t} \right)_{M_1} + \left( \frac{\partial \ln R_{\rm L}}{\partial \ln J} \right)_{M_1,M} \left( \frac{\partial \ln J}{\partial t} \right)_{M_1}
\label{dlnRLdt}
\end{equation}
and
\begin{eqnarray}
\left( \frac{\partial \ln R_{\rm L}}{\partial \ln M_1} \right)_t & \!\!\! = \!\!\! & \left( \frac{\partial \ln R_{\rm L}}{\partial \ln M_1} \right)_{M,J} + \left( \frac{\partial \ln R_{\rm L}}{\partial \ln M} \right)_{M_1,J} \left( \frac{\partial \ln M}{\partial \ln M_1} \right)_t \nonumber \\
& & \mbox{} + \left( \frac{\partial \ln R_{\rm L}}{\partial \ln J} \right)_{M_1,M} \left( \frac{\partial \ln J}{\partial \ln M_1} \right)_t ,
\label{dlnRLdM}
\end{eqnarray}
By analogy with Eq.~(\ref{zetaad}) above, we define
\begin{equation}
\zeta_{\rm L} \equiv \left( \frac{\partial \ln R_{\rm L}}{\partial \ln M_{1}} \right)_t \ .
\label{zetaL}
\end{equation}
Combining Eqs.~(\ref{dlnR1}) and~(\ref{dlnRL}), we obtain
\begin{equation}
{\rm d} \ln {R_1/R_{\rm L}} = \left[ \left( \frac{\partial \ln R_1}{\partial t} \right)_{M_1} - \left( \frac{\partial \ln R_{\rm L}}{\partial t} \right)_{M_1} \right] {\rm d}t + \left( \zeta_{\rm ad} - \zeta_{\rm L} \right) {\rm d} \ln M_1 \ .
\label{dlnDelR}
\end{equation}

Now suppose that the donor star exactly fills its Roche lobe. In order for mass transfer to occur, $R_1$ must increase faster (or decrease slower) than $R_{\rm L}$ would if it were not for mass transfer; that is, the coefficient of ${\rm d}t$ in Eq.~(\ref{dlnDelR}) must be positive. If the mass transfer rate is self-limiting, the stellar radius will parallel the Roche lobe radius as mass transfer proceeds, i.e., ${\rm d} \ln (R_1/R_{\rm L}) \approx 0$. This condition then requires that the second term on the right-hand side of Eq.~(\ref{dlnDelR}) be negative. Since ${\rm d} \ln M_1 < 0$, it follows that self-limiting (stable) mass transfer can only occur if $\zeta_{\rm ad} > \zeta_{\rm L}$. In that case, we can solve directly for the mass transfer rate:
\begin{equation}
\dot{M}_1 = - \frac{M_1}{\zeta_{\rm ad} - \zeta_{\rm L}} \left[ \left( \frac{\partial \ln R_1}{\partial t} \right)_{M_1} - \left( \frac{\partial \ln R_{\rm L}}{\partial t} \right)_{M_1} \right] \ .
\label{zetaad_zetalobe}
\end{equation}
Typically, $\zeta_{\rm ad}$ and $\zeta_{\rm L}$ are of order unity, and mass transfer is driven on a time scale $R_1/\dot{R}_1$. If the donor remains in thermal equilibrium, mass transfer proceeds on a nuclear time scale. If, on the other hand, the donor has been driven out of thermal equilibrium by mass loss, it typically expands much more rapidly, on its thermal time scale, and accordingly mass transfer proceeds on that thermal time scale. In some circumstances, such as among the shorter-period cataclysmic variables or double-degenerate binaries, the driving time scale may be that characterizing contraction of the Roche lobe by angular momentum losses.

Dynamical time scale mass transfer occurs if $\zeta_{\rm ad} \le \zeta_{\rm L}$. Both right-hand terms in Eq.~(\ref{dlnDelR}) are then positive, and the donor star radius is then driven by simple hydrostatic relaxation far beyond its Roche lobe radius. In this case, the mass transfer rate can in principle accelerate up to a truly dynamical rate (cf. Eq.~(\ref{Mdot})). Three-dimensional simulations of this process \citep{rasi96,rick06,rick08} show that the donor envelope is tidally disrupted, engulfing the companion star while tidal torques from the lagging, distorted envelope plunge that companion and the core of the donor into orbits deeply embedded in that envelope. The binary enters common envelope evolution \citep{pacz76}. The common envelope itself lags in rotation behind the embedded cores, which spiral inward as they lose orbital energy and angular momentum to the envelope. Ultimately, either the dissipation of orbital energy suffices to eject the common envelope, revealing the core of the original donor and its companion star\footnote{Because the common envelope has a much higher specific entropy than the surface of the accreting companion, it is much hotter and more tenuous at that interface. The companion can accrete very little matter during common envelope evolution -- see \citet{hjel91}.}, or the cores merge within the common envelope. Examples of the products of common envelope evolution include planetary nebulae with close binary nuclei, cataclysmic variables, and other short-period binaries containing at least one degenerate component. It is important to recognize that common envelope evolution must be very rapid, lest radiative losses quench common envelope ejection, and that energy sources other than orbital energy dissipation are unlikely to contribute significantly to the energy budget of the system \citep{webb07}. Adiabatic mass loss sequences can therefore define the energy difference between initial and final states. This energy difference must be supplied by orbital energy dissipation\footnote{common envelope systems must, of course, also observe angular momentum conservation \citep[cf.][]{nele00,nele01,nele05}, but this is almost always a less stringent constraint on the final system.}.

Useful insights into the behavior of donor stars under adiabatic mass loss can be gleaned from very simplified models. \citet{hjel87} investigated the properties of polytropic models with power-law equations of state, including complete polytropes, composite polytropes ($n = 3$ cores with $n = \frac{3}{2}$ envelopes and $\gamma = \frac{5}{3}$ equations of state), and condensed polytropes ($n = \frac{3}{2}$ envelopes with point mass cores). These models are relevant for some main-sequence stars and giant-branch stars, and illuminate the qualitative (and, within limits, quantitative) behavior of those stars. They were further employed by \citet{sob97} to explore stability criteria for binary mass transfer. However, these models leave many important evolutionary stages of a donor unexplored, and they leave considerable room for improvement even where they may be applicable \citep{egg89,han02,pods02,chen08}. A more sophisticated approach, using realistic stellar models, is both desirable and possible, as shown by \citet{hjel89a,hjel89b}. This paper represents a step toward filling this need.

We describe here a method of construction of stellar adiabatic mass loss sequences. In a given sequence, the initial model corresponds to a full stellar model at an arbitrary state in its evolution. As in the pioneering work of \citet{hjel89a,hjel89b}, mass is removed from the surface while stellar interior at each point relaxes adiabatically to a new state of hydrostatic equilibrium. In Section 2, we describe the modifications to the normal set of stellar structure equations that are needed to construct adiabatic mass loss sequences; how one can recover the internal luminosity profile, and hence thermal relaxation rate, from the adiabatic sequences; and how we have implemented these concepts in a numerical code. Adiabatic mass loss sequences constructed in this way yield $\zeta_{\rm ad}$, fixing the critical mass ratio for dynamical time scale mass transfer, but they also reveal thresholds for a delayed dynamical instability arising when the donor is stripped down to a nearly isentropic core. These applications are described in Section 3. In Section 4, we discuss the calculation of the total energy of a donor star as it loses mass adiabatically, and the application of those results to determining in what orbit, if any, a binary can survive common envelope evolution. We conclude with a brief appraisal of the relevance of our results to population synthesis studies in Section 5.

\section{Adiabatic Mass Loss}
\label{adiab}

Adiabatic mass loss models rest on two basic approximations: that in the limit of rapid mass loss, the response of the donor star is everywhere locally adiabatic; and that hydrostatic equilibrium nevertheless prevails throughout the bulk of the interior of the donor star. The first of these approximations implies not only that heat flow can be ignored, but that (under the Schwarzschild criterion for convection) convective boundaries are preserved with respect to mass. Both the specific entropy profile and the composition profile of the donor interior are therefore fixed along an adiabatic mass loss sequence. The second approximation implies that fluid flow velocities are negligible (highly subsonic) through the interior of the donor. Solutions for the outflow of gas from a lobe-filling star show that this assumption breaks down only in the vicinity of the inner Lagrangian point (see, e.g., \citet{egg06}).

To illustrate the strength and limitations of the adiabatic approximation, we show in Fig.~\ref{entropy-rlof} the specific entropy profiles of the donor star at five epochs during mass transfer for a binary with initial mass $M_{\rm 1i}$ = $2 \,M_{\odot}$, initial mass ratio (donor/accretor) $q_{\rm i}$ = $3.1$, and initial orbital period $P_{\rm i}$ = $0.4 \,\rm {d}$. As mass loss progresses from epoch $a$ to epoch $e$, the mass transfer rate rises from about $10^{-7}\,M_\odot\,{\rm yr^{-1}}$ (a thermal timescale) to about $10^{-5}\,M_\odot\,{\rm yr^{-1}}$ (a quasi-dynamical timescale). We see that the entropy profile of the donor remains nearly unchanged during mass transfer on quasi-dynamical time scales (from curve $c$ to curve $e$), but that even for the thermal timescale mass transfer, thermal relaxation is largely concentrated in the outermost layers, which are immediately stripped away. Little heat transport occurs in the deep interior. This pattern of thermal relaxation typifies stars with radiative envelopes, with rising entropy driven by energy absorption in the outer envelope (as opacity rises rapidly with decreasing temperature during decompression), accompanied by falling entropy in the deep interior (as thermal energy generation supplants nuclear sources as they are frozen out by decompression). Among stars with deep surface convection zones, the entire convective envelope slowly declines in specific entropy, as radiative energy losses from the surface exceed the input of energy at the base of the convection zone (these being choked off by the local rise in opacity as the underlying radiative zone undergoes decompression). Thus, the adiabatic approximation not only defines the asymptotic response of a donor star in the limit of faster-than-thermal time scale mass transfer, but serves as a good approximation to the deep interior response even during thermal time scale mass transfer.

\subsection{The Equations of Stellar Structure}
\label{sub1}

In principle, we can draw the initial model for an adiabatic mass loss sequence from any point in the evolution of the donor star. Following common practice in close binary evolutionary models, we neglect rotational and tidal effects on the structure of that star, and treat it as spherically symmetric. The initial model then satisfies the usual set of four first-order ordinary differential equations of stellar structure, where four structure variables, i.e., pressure $P$, radius $r$, temperature $T$ and luminosity $L$, vary with an independent variable, mass $m$,
\begin{equation}
\frac{{\rm d} \ln P}{{\rm d} m}= - \frac{G m}{4 \pi r^4 P} \ ,
 \label{pressure}
\end{equation}
\begin{equation}
\frac{{\rm d} \ln r}{{\rm d} m}= \frac{1}{4 \pi r^3 \rho}\ ,
\label{radius}
\end{equation}
\begin{equation}
\frac{{\rm d} \ln T}{{\rm d} m} = \frac{{\rm d} \ln P}{{\rm d} m} \ \nabla \ ,  \nonumber
\label{temperature}
\end{equation}
\begin{equation}
\rm {with} \quad \nabla  =   \left\{
\begin{array}{ll}\nabla_{\rm r} \equiv \frac{\textstyle 3 \kappa P L}{\textstyle 16 \pi a c G m T^4} & \rm {Radiative\ zone} \\
\nabla_{\rm conv}   & \rm {Convective\ zone}
\end{array} \right.\ ,
\label{gradient}
\end{equation}
\begin{equation}
\frac{{\rm d} L}{{\rm d} m}= \epsilon - \epsilon_\nu - C_P T \left( \frac{{\rm d} \ln T}{{\rm d} t} - \nabla_{\rm a} \frac{{\rm d} \ln P}{{\rm d} t}\right).
\label{luminosity}
\end{equation}
Here density $\rho$, opacity $\kappa$, adiabatic temperature gradient $\nabla_{\rm {a}} \equiv \left( \partial\ln T/\partial \ln P \right)_s $, nuclear energy generation rate $\epsilon$, neutrino loss rate $\epsilon_{\rm{\nu}}$, and specific heat at constant pressure $C_P = \left( \partial s/\partial \ln T \right)_P $ are functions of $P$, $T$, and the abundances of various nuclear species $X_{i}$. Subscript $s$ means entropy. $G$, $c$ and $a$ are of course the universal gravitational constant, the speed of light and the radiation constant, respectively. $\nabla_{\rm {r}}$ is temperature gradient that would prevail in the absence of convection, while the temperature gradient in a convective zone can be written in a common mixing-length treatment \citep{egg83b} as
\begin{equation}
\nabla_{\rm conv} = \nabla_{\rm r} - \frac{H_P v_{\rm conv}^3}{\alpha \nabla_{\rm a} \chi C_P T}\ .
\label{gradconv}
\end{equation}
The characteristic convective velocity $v_{\rm conv}$ is the real root of the cubic equation
\begin{equation}
\frac{\alpha H_P}{\chi} v_{\rm conv}^3 + v_{\rm conv}^2 + \frac{\chi}{\alpha H_P} v_{\rm conv} = \alpha^2 \nabla_{\rm a} C_P T (\nabla_{\rm r}-\nabla_{\rm a})\ ,
\label{vconv}
\end{equation}
where $H_P$ is the pressure scale height, $\alpha$ the dimensionless mixing-length parameter, and
\begin{equation}
\chi = \frac{4acT^3}{3\kappa \rho^2 C_P}\ .
\label{chi}
\end{equation}
Four boundary conditions are required to close this set of equations. At the center ($m = 0$),
\begin{equation}
r = 0\ ,
\label{rctr}
\end{equation}
\begin{equation}
L = 0\ ;
\label{Lctr}
\end{equation}
and at the surface ($m = M$),
\begin{equation}
L = 4\pi r^2 \sigma T^4\ ,
\label{Lsurf}
\end{equation}
\begin{equation}
\frac{\kappa \left( P_{\rm gas} + \frac{1}{2} P_{\rm rad} \right)}{g} = \frac{2}{3}\ ,
\label{bc_surf}
\end{equation}
where the latter surface boundary condition approximates a classical gray atmosphere. The four equations are solved for a given distribution of $X_{{i}}(m)$, which is updated according to nuclear reactions and convective mixing. Stellar mass loss can be included by applying the surface boundary conditions at a moving surface such that $dM = \dot{M}\,{\rm d}t$. Roche lobe overflow in a binary system can likewise be accommodated either by expressing $\dot{M}$ explicitly in terms of the degree of Roche lobe overflow (see, for example, the Appendix to this paper), or (in the approximation that $R = R_{\rm L}$) by rewriting the structure equations above in terms of $r$ rather than $m$ as the independent variable, and applying the surface boundary conditions at $r = R_{\rm L}(m)$.

In adiabatic mass loss models, the equations of energy transport and conservation (Eqs.~(\ref{temperature}) and~(\ref{luminosity}), respectively) are supplanted by algebraic conditions fixing the specific entropy and composition profiles:
\begin{equation}
s(m) = s_0(m)
\label{entropy}
\end{equation}
\begin{equation}
X_{i}(m) = X_{i,0}(m)\ ,
\label{composition}
\end{equation}
where the zero subscripts denote the values of these quantities at the same mass coordinate in the initial model of the sequence. Eqs.~(\ref{pressure}),~(\ref{radius}),~(\ref{entropy})and~(\ref{composition}), together with appropriate boundary conditions (identified below), are sufficient to describe completely the state of the gas at any point in the interior of the donor. The local temperature, $T = T(P,s,X_{i})$, follows implicitly.

To understand the internal relaxation processes set in motion by adiabatic mass loss, we would also like to recover the luminosity profile through the interior of the donor star. This we can accomplish by writing the local energy flux through the stellar interior in terms of the local entropy gradient, which is a known quantity. Given the local state of the gas, the entropy gradient can be written in terms of the pressure and composition gradients:
\begin{eqnarray}
\frac{{\rm d}s}{{\rm d}m}  & \!\!\! = \!\!\! & \left( \frac{\partial s}{\partial P} \right)_{T,X_i} \frac{{\rm d}P}{{\rm d}m} + \left( \frac{\partial s}{\partial T} \right)_{P,X_i} \frac{{\rm d}T}{{\rm d}m} + \sum_i \left( \frac{\partial s}{\partial X_i} \right)_{P,T} \frac{{\rm d}X_i}{{\rm d}m} \nonumber \\
  & \!\!\! = \!\!\! & \left[ \left( \frac{\partial s}{\partial \ln P} \right)_{T,X_i} + \left( \frac{\partial s}{\partial \ln T} \right)_{P,X_i} \frac{{\rm d} \ln T}{{\rm d} \ln P} \right] \frac{{\rm d} \ln P}{{\rm d}m} + \sum_i \left( \frac{\partial s}{\partial X_i}\right)_{P,T}\frac{{\rm d}X_i}{{\rm d}m} \nonumber \\
  & \!\!\! = \!\!\! & C_P \left[ \nabla + \left( \frac{\partial s}{\partial \ln P} \right)_{T,X_i} \left( \frac{\partial s}{\partial \ln T} \right)_{P,X_i}^{-1} \right] \frac{{\rm d}\ln P}{{\rm d}m} + \sum_i \left( \frac{\partial s}{\partial X_i} \right)_{P,T} \frac{{\rm d}X_i}{{\rm d}m} \nonumber \\
  & \!\!\! = \!\!\! & C_P (\nabla-\nabla_{\rm a})\ \frac{{\rm d}\ln P}{{\rm d}m} + \sum_i \left( \frac{\partial s}{\partial X_i} \right)_{P,T} \frac{{\rm d}X_i}{{\rm d}m}\ ,
\label{dsdm}
\end{eqnarray}
where we have invoked the definitions of specific heat, $C_P = (\partial s/\partial \ln T)_{P,X_i}$, and of the adiabatic temperature gradient, $\nabla_{\rm a} = (\partial \ln T/\partial \ln P)_{s,X_i} = - (\partial s/\partial \ln P)_{T,X_i}(\partial s/\partial \ln T)_{P,X_i})^{-1}$. As overlying layers are stripped away, successively deeper mass layers are brought toward the stellar surface, pressure in those layers decreases, and $|{\rm d \ln} P/{\rm d}m|$ increases accordingly. The ambient temperature gradient, $\nabla$, is therefore driven toward the adiabatic temperature gradient, $\nabla_{\rm a}$, since ${\rm d}s/{\rm d}m$ and ${\rm d}X_{i}/{\rm d}m$ are fixed in Equation~(\ref{dsdm}).

Let us define an $\it {effective}$ entropy gradient, ${\rm d} \tilde{s}/ {\rm d} m$, as
\begin{equation}
\frac{{\rm d}\tilde{s}}{{\rm d}m} \equiv \frac{{\rm d}s}{{\rm d}m} - \sum_{i} \left( \frac{\partial s}{\partial X_{i}} \right)_{P,T} \frac{{\rm d}X_{i}}{{\rm d}m}\ = C_P (\nabla-\nabla_{\rm a})\ \frac{{\rm d}\ln P}{{\rm d}m} ,
\label{seff}
\end{equation}
Then the ambient dimensionless temperature gradient, $\nabla$, can be written as
\begin{equation}
\nabla = \nabla_{\rm a} + \frac{{\rm d}\tilde{s}}{{\rm d}m} \frac{1}{C_P} \left( \frac{{\rm d} \ln P}{{\rm d} m} \right)^{-1} = \nabla_{\rm a} - \frac{4\pi r^4 P}{GmC_P} \frac{{\rm d} \tilde{s}}{{\rm d} m}\ .
\label{grad}
\end{equation}
The value of ${\rm d} \tilde{s}/ {\rm d} m$ can readily be obtained at each layer of an adiabatic mass loss model, and so $\nabla$ is a known quantity. If $ {\rm d} \tilde{s}/ {\rm d} m \ge 0$, then $\nabla \le \nabla_{\rm a}$; such a layer is stable against convection. In this case, $\nabla_{\rm r} = \nabla$, and
\begin{equation}
L = L_{\rm r} = \frac{16 \pi a c G m  T^4}{3 \kappa P} \nabla .
\label{Lrad}
\end{equation}
If ${\rm d} \tilde{s} / {\rm d}m < 0$, on the other hand, then $\nabla > \nabla_{\rm a}$. Such a layer is convectively unstable and energy is transported by both radiation and convection. The temperature gradient that would prevail in the absence of convection is (cf. Eqs.~\ref{gradconv},~\ref{vconv}, and~\ref{chi}):
\begin{equation}
\nabla_{\rm r} = \nabla + \frac{4 \chi^2}{27 H_P^2 \alpha^4 C_P T \nabla_{\rm a}}\,\omega^3 ,
\label{gradr}
\end{equation}
where
\begin{equation}
\omega = \frac{27}{4} \left( \sqrt{1+\frac{8A}{2187}} -1  \right) ,
\label{omegaconv}
\end{equation}
and
\begin{equation}
A = \left( \frac{{\rm d} \tilde{s}/{\rm d}m}{C_P \ {\rm d ln}P/ {\rm d}m } \right)
\left( \frac{27 H_P^{2} \alpha^4 C_P T \nabla_{\rm a}}{4 \chi^2}\right)\ .
\label{Aconv}
\end{equation}
The total luminosity is then
\begin{equation}
L = \frac{16 \pi a c G m  T^4}{3 \kappa P} \nabla_{\rm r}\ ,
\label{rad_diff}
\end{equation}
where $\nabla_{\rm r}$ is calculated from Eqs.~(\ref{gradr}),~(\ref{omegaconv}) and~(\ref{Aconv}).
Thus the additional information needed to reconstruct the luminosity profile in an adiabatic model is already carried in the entropy and composition gradients.

It remains to specify the boundary conditions for the adiabatic mass loss models. At the center of the star ($m = 0$), clearly
\begin{equation}
r = 0
\end{equation}
as before. At the surface of the star, however, there is a fundamental inconsistency between the assumption of adiabatic relaxation (implying no radiative losses) and the finite effective temperature that finite photospheric optical depth inevitably demands. In the outer atmosphere of a star, at small optical depth, radiative relaxation always proceeds more rapidly than dynamical relaxation, so that purely adiabatic relaxation is impossible. However, at large optical depth, the diffusion approximation remains valid, and it fixes the stellar luminosity in terms of the ambient entropy gradient, as we have seen above. We therefore resolve this inconsistency by insisting on continuity of luminosity at the stellar surface. Substituting the blackbody law into the radiative diffusion equation (Eq. (\ref{rad_diff})), we obtain for the surface boundary condition
\begin{equation}
\nabla_{{\rm r}} = \frac{3}{16} \frac{P \kappa }{g}\ .
\label{pts}
\end{equation}
This condition is tantamount to one relating the entropy gradient to the entropy itself at the photosphere.

\subsection{Numerical Implementation}
\label{subsecNCC}

The numerical code to solve the equations of the adiabatic mass loss model in section~\ref{sub1} is written in $FORTRAN95$, based on the stellar evolution code developed by \citet{egg71, egg72, egg73} and \citet{pax04}. This code has been widely used and is updated with the latest input physics, as described by \citet{han94, han03} and \citet{pols95, pols98}. The essential features of this code have been retained in our adiabatic mass loss model, namely (a)~the use of an adaptive, moving, non-Lagrangian mesh; (b)~the treatment of both convective and semiconvective mixing as diffusion processes; and (c)~the simultaneous, implicit solution of both the stellar structure equations and (in the evolutionary models) the chemical composition equations, including convective mixing.

Our code uses an equation of state, described by \citet{eff73} and \citet{pols95}, that includes pressure ionization, Coulomb interactions and dissociation of molecular hydrogen. The opacity tables are taken from \citet{af94a, af94b} in the low-temperature limit (where molecules become important), from \citet{itoh83} in the high-density, high-temperature regime (where electron conduction becomes important), and from \citet{rog92} elsewhere. The nuclear reaction rates are interpolated in temperature, mostly from \citet{cau88} and \citet{cau85}. Electron screening theory, including strong and intermediate screening, comes from \citet{grab73}. The neutrino energy loss rates resulting from pair, photo-, plasma, bremsstrahlung, and recombination neutrino processes originate from \citet{itoh96}.

The adaptive, non-Lagrangian mesh automatically distributes mesh points at uniform intervals in a function, $Q$, of the local structure variables \citep{egg71}. The choice of function $Q$ is in principle arbitrary, subject only to the constraint that it vary monotonically from center to surface of the stellar model. Its introduction involves rewriting the stellar structure equations in terms of $k$, the mesh point number, as the independent (radial) variable, and introducing an additional `structure' equation for the mesh function:
\begin{equation}
\frac{{\rm d}Q}{{\rm d}k} = \frac{Q(N) - Q(1)}{N - 1} = \mbox{constant}\ ,
\label{dQdk}
\end{equation}
where $N$ is the number of mesh points in the model. In our case, we employ 400 mesh points, distributed according to the mesh function
\begin{eqnarray}
Q & \!\!\! = \!\!\! & 0.05 \ln (P/{\rm dyn~cm^{-2}}) + 0.15 \ln \left[ (P/{\rm dyn~cm^{-2}}) + 10^{15} \right] + 0.45 \ln \left[ T/(T+2 \cdot 10^4{\rm K}) \right] \nonumber \\
& & + \ln \left[ 0.02\,m_{\rm c}^{2/3}/(0.02\,m_{\rm c}^{2/3} + m^{2/3}) \right] - 0.05 \ln \left[ 1 +  (r/10^9\ {\rm cm})^2 \right]\ ,
\label{Qmesh}
\end{eqnarray}
where $m_{\rm c} = 3.5 \rho ({P/G\rho^2})^{3/2}$. We find that his mesh function provides good resolution of structural features of interest throughout our evolutionary and mass loss sequences. Taking the derivative of Eq.~(\ref{Qmesh}), and substituting Eqs.~(\ref{pressure},~\ref{radius}, and~\ref{temperature}), with $\nabla$ determined from Eq.~(\ref{grad}) in lieu of Eq.~(\ref{gradient}), we can find ${\rm d} Q/{\rm d} m$. Using Eq.~(\ref{dQdk}), we can then cast the structure equations in terms of $k$ rather than $m$ as the independent (radial) variable.

In practice, we use $\ln f$ (a degeneracy parameter related to the electron chemical potential --- see \citet{eff73, egg06}), and $\ln (T/{\rm K})$ as our state variables, and $\ln (r/10^{11}\ {\rm cm})$, $(m/10^{33}\ {\rm g})$, and $Q$ as our global variables. The differential equations for $P(k)$, $r(k)$, $m(k)$, and $Q(k)$ are written in difference form, as usual, but we include also the algebraic equation $\tilde{s}_0(m(k)) = \tilde{s}(f,T,X_{i})$ in the solution matrix as a convenient tool for satisfying the entropy constraint. The two additional differential equations require boundary conditions. At the surface ($k = 1$),
\begin{equation}
\nabla_{{\rm r}} = \frac{3}{16} \frac{P \kappa }{g}\ ,
\end{equation}
\begin{equation}
m = M\ .
\label{msurf}
\end{equation}
At the center ($k = N$), we should have $r = 0$, as before,  and $m = 0$, but to avoid singular behavior in that limit, the central mesh point is offset:
\begin{equation}
m = -\frac{{\rm d} m}{{\rm d} k}\ ,
\label{mK}
\end{equation}
\begin{equation}
r = -\frac{{\rm d} r}{{\rm d} k}\ .
\label{rK}
\end{equation}

As intimated above, the construction of adiabatic mass loss sequences involves the interpolation in this adaptive mesh of  the specific entropy $s$ at the corresponding mass coordinate in the initial model of the sequence. Near the surface of an initial model, $s$ typically varies very rapidly with mass, while the mass increment from one mesh point to the next becomes a very small fraction of the stellar mass. Because of the limitations of numerical accuracy (we typically require that the \emph{average} correction to structure variables in a model deemed `converged' not exceed one part in $10^7$), the situation arises frequently that the nominal error in mass coordinate $m$ is comparable with (or even exceeds) the mass difference between one mesh point and the next. In this case, the nominal mass coordinates of mesh points near the surface may not be well-ordered, making the accurate interpolation in mass of the specific entropy and its gradient impossible. We avoid this problem altogether by using not the mass coordinate $m$, but the mass complement, $\overline{m} \equiv m - M$, where $M$ is the total mass of the star, over the outer half of our mesh. Our convergence criterion, now applied to $\overline{m}$, ensures a mesh well-ordered in mass. For similar reasons of numerical accuracy, we tabulate ${\rm d} \tilde{s}/{\rm d} m$, derived from the local stellar luminosity as described above, for the initial model of each adiabatic sequence, and interpolate in this gradient, rather than numerically differentiating $s$ and $X$ themselves, in reconstructing the internal luminosity profiles of models in our adiabatic mass loss sequences.

\begin{figure}
\begin{center}
\resizebox{16cm}{!}{\rotatebox{-90}{\includegraphics{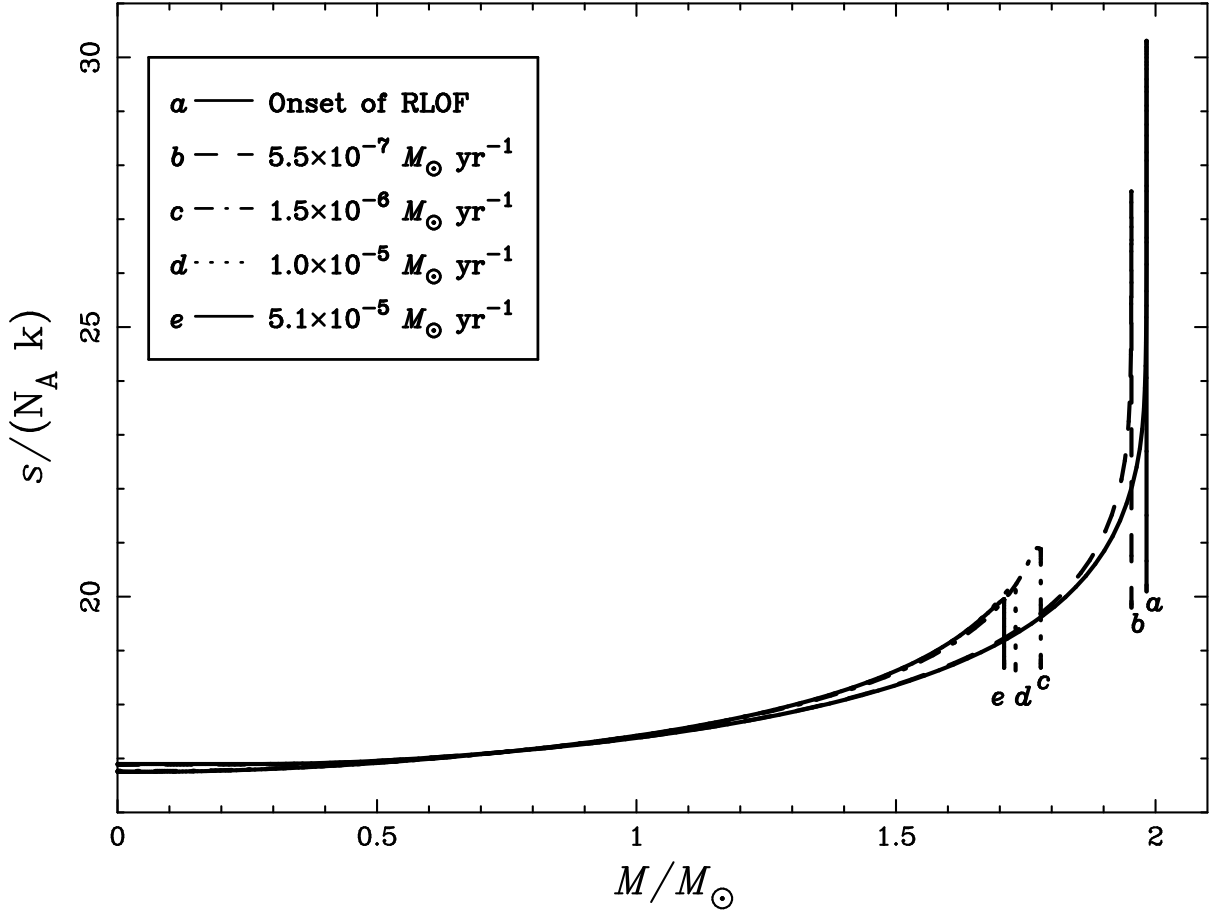}}}
\caption{Specific entropy profiles at five epochs in a $2.0\ M_{\odot}$ donor undergoing conservative Roche lobe overflow (RLOF) to a $0.65\ M_{\odot}$ companion. Curve $a$ marks the beginning of RLOF. The mass-transfer reaches rate quickly rises from $5.5\times10^{-7}\,M_\odot {\rm yr}^{-1}$, roughly a thermal time scale rate (curve $b$), to $5.1\times10^{-5}\,M_\odot {\rm yr}^{-1}$ (curve $e$).}
\label{entropy-rlof}
\end{center}
\end{figure}

\begin{figure}
\begin{center}
\resizebox{!}{16cm}{\rotatebox{0}{\includegraphics{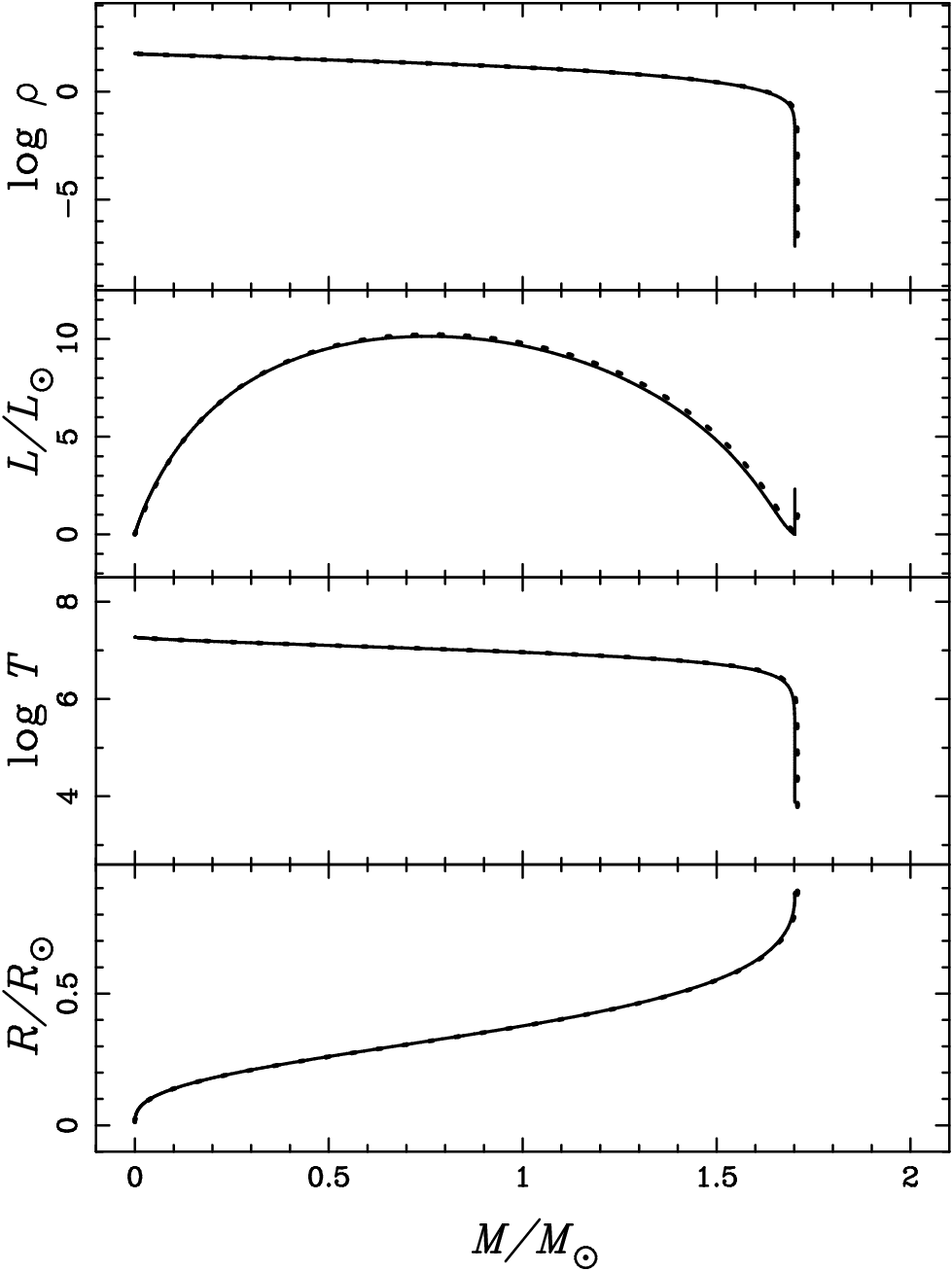}}}
\caption{Density (${\rm g \ cm^{-3}}$), luminosity, temperature (K) and radius profiles for the $2.0\  M_{\odot}$ donor shown in Fig.~\ref{entropy-rlof} at epochs $c$ (solid lines) and $e$ (dotted lines).}
\label{rtlrho_m}
\end{center}
\end{figure}

As a demonstration of the ability of our adiabatic mass loss code to reproduce faithfully the response of a donor star undergoing rapid mass loss, we have calculated an adiabatic mass loss sequence, beginning at model ($c$) of the time-dependent mass transfer calculation shown in Fig.~\ref{entropy-rlof}. At this point, the mass loss rate from the donor has grown to roughly a thermal time scale rate. In Fig.~\ref{rtlrho_m} we show comparisons of the interior density, luminosity, temperature, and radius profiles in the adiabatic models with those in the time-dependent calculation when both models have reached point ($e$) of Fig.~\ref{entropy-rlof}. Only very minor differences appear in layers very close to the surface, but in general these two calculations are in excellent agreement, giving strong support to the use of adiabatic mass loss sequences to model rapid mass transfer.

\section{Thresholds for  Dynamical Instability}
\label{secresults}

\subsection{Physical Significance of $\zeta_{\rm ad}$}
\label{significance}

An important application of adiabatic mass loss sequences is the evaluation of threshold conditions for the onset of dynamical time scale mass transfer. As discussed in Section~\ref{intro}, this condition corresponds formally to $\zeta_{\rm ad} \le \zeta_{\rm L}$; in that case, a sufficiently rapid perturbation to the donor star (i.e., one faster than a thermal time scale) can lead to unbridled growth of the donor radius beyond its Roche lobe (cf. Eq.~\ref{dlnDelR}). On the other hand, if $\zeta_{\rm ad} > \zeta_{\rm L}$ the donor will transfer mass on a slower time scale, on a nuclear or angular momentum loss time scale if the donor thermal equilibrium radius-mass relation ($\zeta_{\rm eq} \equiv (\partial \ln R/\partial \ln M)_{\rm eq}$) is sufficiently steep ($\zeta_{\rm eq} > \zeta_{\rm L}$), or on a thermal time scale if thermal relaxation tends to drive the donor beyond its Roche lobe ($\zeta_{\rm eq} < \zeta_{\rm L} < \zeta_{\rm ad}$).

\begin{figure}
\begin{center}
\resizebox{!}{16cm}{\rotatebox{0}{\includegraphics{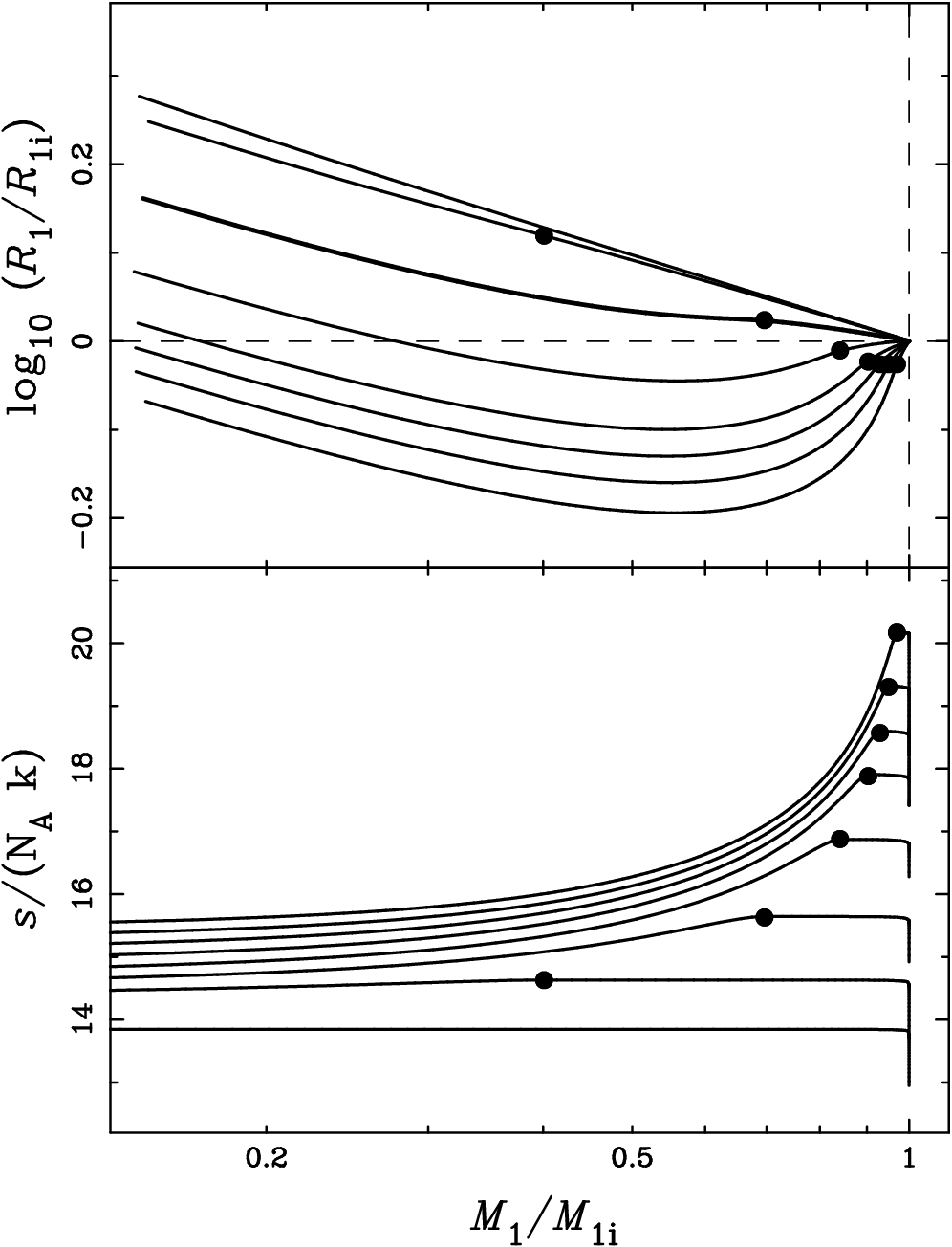}}}
\caption{Adiabatic response curves (top) and the entropy profiles (bottom) for low-mass ZAMS models, of initial mass $M_{\rm 1i}$ = 0.30, 0.40, $\dots$, 0.90, 1.00 $M_{\odot}$. Solid  circles mark the boundaries between the radiative core and the convective envelope. Initial masses increase from top to bottom in the radius response curves (upper panel), and from bottom to top in the entropy profiles (bottom panel).}
\label{newrmsm}
\end{center}
\end{figure}

Adiabatic responses of stars to mass loss depend strongly on the variation of specific entropy with mass in their outermost layers.  This behavior is illustrated in Fig.~\ref{newrmsm} for lower main sequence stars.  As one progresses down the main sequence, the surface convection zone grows deeper, and, as illustrated here, the initial contraction of stellar radius with decreasing mass becomes less and less severe.  In the limit of fully convective stars ($M_{\rm i} = 0.30 M_{\odot}$ in Fig.~\ref{newrmsm}), their behavior approximates the $R/R_{\rm 1i} \sim (M/M_{\rm 1i})^{-1/3}$ dependence that follows from the homologous expansion of an isentropic gas sphere supported by classical ideal gas pressure.  In contrast, stars on the upper main sequence, with nearly completely radiative envelopes, undergo very rapid contraction with adiabatic mass loss, since their high-entropy outer layers occupy a disproportionately large fraction of their stellar volume.  Broadly speaking, then, stars with convective envelopes tend toward dynamically unstable mass transfer, while those with radiative envelopes tend toward stability.

The situation in reality is somewhat more complex.  As is evident in Fig.~\ref{newrmsm}, the radius-mass exponents $\zeta_{\rm ad}$, $\zeta_{\rm eq}$, and $\zeta_{\rm L}$ are not constant over the course of mass transfer, but vary continuously, in some cases very rapidly.  This is true particularly of $\zeta_{\rm ad}$ in the case of a star with a radiative envelope as it first fills its Roche lobe.  Typically, the specific entropy rises extremely rapidly through the surface layers of such a star (see Fig.~\ref{entropy-rlof}), which therefore decrease rapidly in density.  Loss of even a small amount of mass leads to rapid reduction in the stellar radius, so $\zeta_{\rm ad}$ is initially very large and positive.  Donor stars with radiative envelopes are therefore typically strongly stable against dynamical time scale mass transfer, at least initially.  However, as mass loss proceeds, and the surface layers are stripped away, entropy gradients flatten, and the initially rapid stellar contraction moderates considerably.  If the deep interior lacks strong entropy gradients (a circumstance typically prevailing during core hydrogen burning), then as the donor star is stripped down to this region it may begin expanding with further mass loss, much as an (isentropic) convective star expands with decreasing mass.  Under these circumstances, the donor star may belatedly become unstable to \emph{delayed dynamical mass transfer}.

\subsection{The Delayed Dynamical Instability}
\label{DDinstab}

\begin{figure}
\begin{center}
\resizebox{16cm}{!}{\rotatebox{-90}{\includegraphics{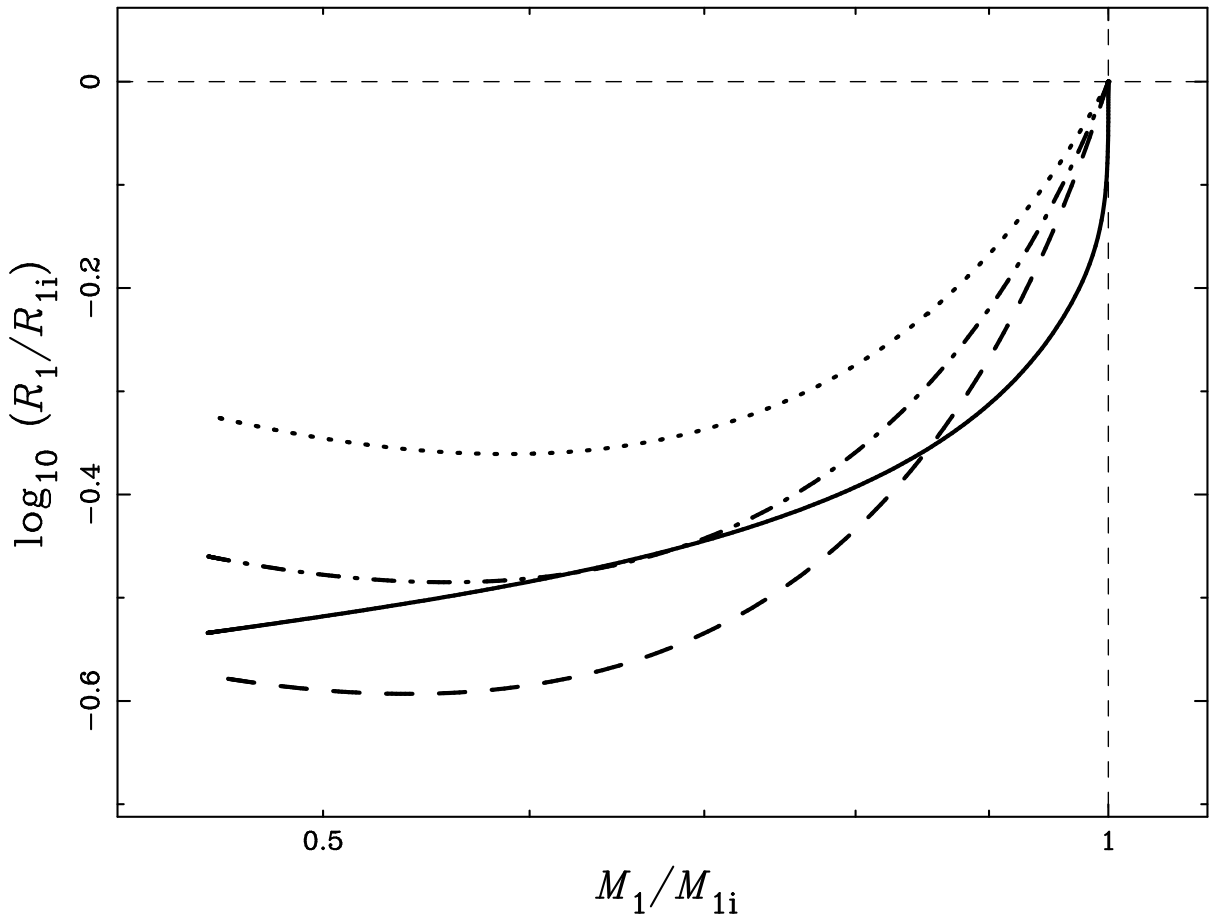}}}
\caption{Radial response curves for a $3.2\ M_{\odot}$ terminal main sequence donor star.  The solid line traces the adiabatic mass loss sequence.  Broken lines mark the Roche lobe radius as a function of mass for three initial mass ratios: $q_0 \equiv M_{\rm 1i}/M_{\rm 2i} = 3.0$ (dotted line), 3.757 (dash-dotted line, and 4.5 (dashed line), all assuming conservative mass transfer.  In the limit that the donor star truly responds to mass loss adiabatically, the donor star becomes unstable to dynamical time scale mass transfer when the adiabatic radius exceeds the Roche lobe radius.  Thus, the $q_0 = 3.0$ binary is stable against dynamical mass transfer throughout mass loss, while the $q_0 = 4.5$ binary reaches dynamical instability when the donor star has been stripped to $M_1/M_{\rm 1i} = 0.853$.  This is a delayed dynamical instability, because it erupts only after an extended period of thermal time scale mass tranfer.  The threshold initial mass ratio for delayed dynamical instability occurs where the Roche lobe radius and adiabatic radius reach tangency, roughly at $M_1/M_{\rm 1i} = 0.653$, $q_0 = 3.757$.}
\label{zeta_ad_3_2msun}
\end{center}
\end{figure}

The adiabatic response of a $3.2\ M_{\odot}$ terminal main sequence star, shown in Fig.~\ref{zeta_ad_3_2msun}, illustrates the nature of this delayed dynamical instability.  This star has a deep radiative envelope, and we see that in the adiabatic limit, its initial response is indeed extremely rapid contraction.  Recalling that even in thermal time scale mass transfer, little thermal relaxation occurs in the deep interior of a donor star (see section~\ref{adiab}, Fig.~\ref{entropy-rlof}), we can regard the adiabatic mass loss sequence as approximating the \emph{minimum} radius of the donor star as a function of its mass.  We see that, in this example, a binary with an initial mass ratio $q_{\rm i} = 3.0$, assumed to transfer mass conservatively, never reaches this minimum radius in the course of mass transfer; in this case, the mass transfer rate need never exceed a thermal rate in order for the donor star to track its Roche lobe through the mass transfer episode.  In contrast, a binary with an initial mass ratio $q_{\rm i} = 4.5$ undergoes such rapid Roche lobe contraction (again in the conservative approximation) that, despite the rapid initial adiabatic contraction of the donor, the Roche lobe radius reaches the minimum radius set by the adiabatic mass loss sequence (intersection of the dashed line with the solid line in Fig.~\ref{zeta_ad_3_2msun}) when the donor has been reduced to a fraction 0.86 of its initial mass.  In this case, the system may transfer mass at a thermal rate up to that point, but the donor is then no longer able to remain within its Roche lobe, and dynamical time scale mass transfer must ensue.  This is the delayed dynamical instability.  At an initial mass ratio $q_{\rm i} = 3.757$ (dash-dotted curve), the Roche lobe radius is tangent to the adiabatic mass-radius relation (at $M/M_{\rm i} = 0.653$), a condition which then marks the threshold for delayed dynamical instability, in the limit that the donor star response is truly adiabatic, and that mass transfer is conservative.

\begin{figure}
\begin{center}
\resizebox{16cm}{!}{\rotatebox{-90}{\includegraphics{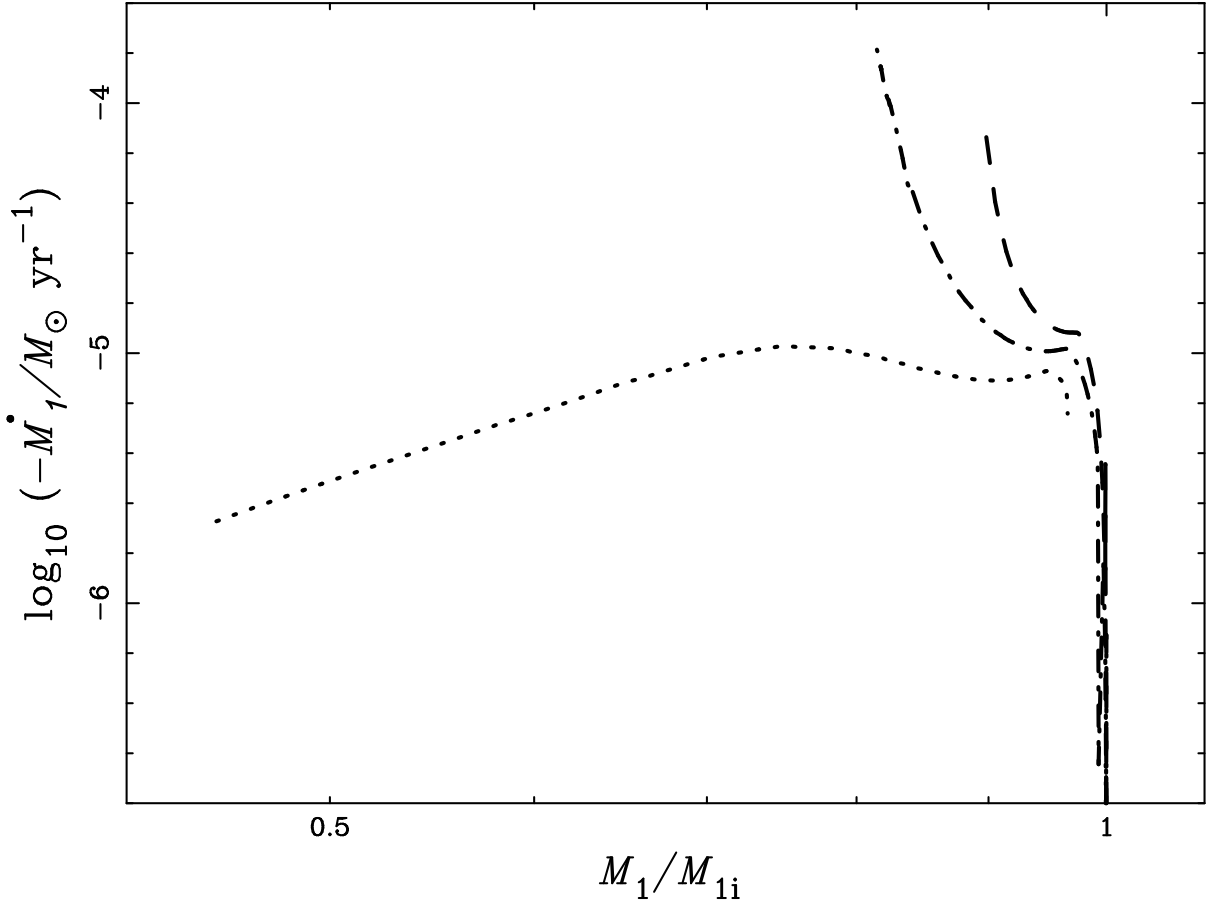}}}
\caption{Mass loss rates of the $3.2 M_{\odot}$ terminal main sequence models of Fig.~\ref{zeta_ad_3_2msun} from fully time-dependent evolutionary calculations of conservative Roche lobe overflow (based on the  method by \citet{han00}). The onset of dynamical time scale mass transfer from these calculations occurs at $M_1/M_{\rm 1i} < 0.905$ for $q_0 = 4.5$ and $M_1/M_{\rm 1i} < 0.814$ for $q_0 = 3.757$.  These calculations stop short of fully dynamical time scale mass transfer, but suggest that the time-dependent models may be slightly more prone to instability.}
\label{mdot_3_2msun}
\end{center}
\end{figure}

Since the delayed dynamical instability follows a period of slower, thermal time scale mass transfer, in which some measure of thermal relaxation inevitably occurs, the question arises, how good is the adiabatic approximation for the threshold of delayed dynamical instability?  We can glean at least a partial answer to this question by comparing the adiabatic results in the example above with a family of time-dependent mass transfer models spanning the same range of initial mass ratios, and again assuming conservative mass transfer, but now with the full set of stellar structure equations.  We again assume conservative mass transfer, but now require the donor star to exactly fill its Roche lobe throughout mass transfer.  The results of this exercise are shown in Fig.~\ref{mdot_3_2msun}, showing mass transfer rates as a function of donor mass on the same scale as Fig.~\ref{zeta_ad_3_2msun} to facilitate comparison.  In this figure, the onset of delayed dynamical instability is marked by a rapidly accelerating mass transfer rate as a function of decreasing donor star mass.  We see that the mass transfer rates in systems with initial mass ratios $q_{\rm i} = 4.5$ and $q_{\rm i} = 3.757$ both indeed accelerate well beyond the thermal time scale rate, while the transfer rate for the system with initial mass ratio $q_{\rm i} = 3.0$ never accelerates beyond a thermal rate.  The mass loss rates for the unstable systems $q_{\rm i} = 4.5$ and $q_{\rm i} = 3.757$ peak at donor star masses $M/M_{\rm i} = 0.905$ and $M/M_{\rm i} = 0.814$, respectively, somewhat earlier than predicted by the adiabatic calculations seen in Fig.~\ref{zeta_ad_3_2msun}, indicating that the modest thermal relaxation that occurs in the time-dependent calculation drives the donor slightly in the direction of instability.  In this case, we may overestimate the critical initial mass ratio for delayed dynamical instability by about $10\%$, which we may then consider a mark of the accuracy and reliability with which we can define this threshold from adiabatic mass loss sequences alone.

\subsection{Superadiabatic Expansion}
\label{superadiabatic}

A different sort of complication arises among luminous giant branch stars, in which convection becomes very inefficient in their outer envelopes.  The ambient density there falls so low that an increasing fraction of the stellar luminosity must be carried by radiation, notwithstanding the fact that the local medium is unstable to convection.  This condition drives the local temperature gradient, $\nabla$, well above the adiabatic gradient, $\nabla_{\rm a}$ (see subsection~\ref{sub1}).  The entropy gradient thus becomes negative over a significant fraction of the outer envelope; in effect, the outer envelope forms a cool, dense blanket, suppressing the higher-entropy convective envelope contained within it.  Removal of this blanketing superadiabatic layer leads to an extremely rapid initial expansion of the star. As seen in Fig.~\ref{GB_expansion}, this \emph{superadiabatic expansion} can be quite dramatic in luminous giant stars, but in fact it is a feature of any donor star with a surface convection zone. Among less luminous stars (generally, those which have not yet reached the giant branch), the mass and radial extent of the superadiabatic layer is small enough that its effect on threshold conditions for sustained dynamical time scale mass transfer are negligible.  However, superadiabatic expansion is a pervasive feature of luminous giant branch stars.

The behavior seen in Fig.~\ref{GB_expansion} reveals certain limitations to the utility of spherically-symmetric adiabatic mass loss models in analyzing the stability of mass transfer.  In reality, mass is not removed uniformly over the whole surface of the donor, but from an initially small region about the inner Lagrangian point, $L_1$.  Moreover, sufficiently close to the stellar photosphere, radiative relaxation inevitably becomes rapid compared with convective turnover time scales, so that conditions for adiabatic outflow cannot be satisfied.  Indeed, for a giant branch star near Eddington luminosities, the thermal time scale of the entire envelope may approach its dynamical time scale.  Without full three-dimensional models, including both advective and radiative energy transport, we cannot be certain how much of this superadiabatic expansion is real, and how much an artifact of our simplifying assumptions.  Nevertheless, there exists a real physical basis for this behavior, and the adiabatic models show that it is at least energetically possible for superadiabatic expansion to trigger mass transfer when the donor photosphere still lies well within its Roche lobe.  In the convective envelope, hydrostatic equilibrium describes an average state of the gas, but in fact it consists of rising and descending flows, which may differ considerably from each other in specific energy or entropy, and approach sonic velocity where the mean entropy gradient is strongly superadiabatic.  That kinetic energy is in principal capable of launching gas through the saddle point in gravitational potential at $L_1$ even as the mean stellar photosphere lies well within the donor star Roche lobe.  Rising convective flows may be captured by the companion star, with the result that mass transfer rates can fluctuate wildly on convective turnover time scales\footnote{Episodic mass transfer driven by superadiabatic expansion was first postulated by Bath (\citet{bath72,bath75}; see also~\citet{pap75}).}.

\subsection{Critical Mass Ratios}
\label{qcrit}

In future papers (Ge et al., in preparation), we will present results of a survey of adiabatic responses of stars throughout the Hertzsprung-Russell diagram.  Inasmuch as one of the principal objectives of this survey is to establish the conditions under which a lobe-filling donor star of arbitrary evolutionary state is stable or unstable to dynamical time scale mass transfer, we describe here the issues which arise in characterizing our results for that purpose.

We idealize mass transfer between binary components as fully conservative, by which we mean specifically that the total mass of the binary
\begin{equation}
M_{\rm tot} = M_1 + M_2
\label{mtot}
\end{equation}
and orbital angular momentum of the binary
\begin{equation}
J_{\rm orb} = \left( \frac{M_1^2 M_2^2}{M_1 + M_2}\,G A \right)^{1/2}
\label{jorb}
\end{equation}
are constant. (We assume, of course, that the orbital eccentricity is throughout negligibly small.) Then, using \citeauthor{egg83a}'s \citeyearpar{egg83a} approximation for the Roche lobe radius,
\begin{equation}
r_{\rm L}(q) = \frac{0.49 q^{2/3}}{0.6 q^{2/3} + \ln(1+q^{1/3})}\ ,
\label{rlobe}
\end{equation}
where the mass ratio $q$ is defined, as before, as the ratio of donor star mass to accretor mass, we can write the Roche lobe mass-radius relation directly in terms of the binary mass ratio:
\begin{equation}
\zeta_{\rm L} \equiv \left(\frac{\partial \ln R_{\rm L}}{\partial \ln M_1}\right)_{J,M} = \left[\frac{2\ln(1+q^{1/3})-q^{1/3}/(1+q^{1/3})}{3[0.6q^{2/3} + \ln(1+q^{1/3})]}-2\left(\frac{1-q}{1+q}\right) \right](1+q)\ .
\label{zetalobe}
\end{equation}
Given $\zeta_{\rm ad}$ from the adiabatic mass loss sequence for a donor star of interest, Eq.~(\ref{zetalobe}) then implicitly defines a corresponding critical mass ratio, $q_{\rm crit}$, satisfying the equation
\begin{equation}
\zeta_{\rm ad} = \zeta_{\rm L}(q_{\rm crit})
\label{qcrit_zeta}
\end{equation}
above which a binary containing that donor star is unstable to dynamical time scale mass transfer.

In reality, mass transfer is never truly conservative, and Eq.~(\ref{zetalobe}) is at best an approximation to the dependence of Roche lobe radius on mass ratio.  In contrast, to the extent that our spherically-symmetric models are appropriate, $\zeta_{\rm ad}$ is intrinsic to the donor star alone, independent of the mass or nature of the companion star. In this sense, it is the more fundamental parameter by which we should characterize our results. However, as noted above, $\zeta_{\rm ad}$ varies continuously over the course of mass loss, and the decrease in $\zeta_{\rm ad}$ with decreasing donor mass is crucial to the existence of a delayed dynamical instability. Conditions for the onset of that instability therefore depend on the extent and nature of mass and angular momentum losses accompanying mass transfer.  For the sake of clarity and simplicity, we here and henceforth implicitly assume conservative mass transfer in quantifying thresholds for dynamical time scale mass transfer, and denote by $q_{\rm crit}$, specifically the threshold \emph{initial} mass ratio for dynamical instability, and define $\zeta_{\rm crit} = \zeta_{\rm L}(q_{\rm crit})$, the corresponding critical donor mass-radius exponent marking this same threshold.

The evaluation of $q_{\rm crit}$ is complicated by the fact that the distinction between a prompt dynamical instability and a delayed dynamical instability is to some extent artificial. Even if the donor star satisfies the inequality for dynamical instability ($\zeta_{\rm ad} < \zeta_{\rm L}$) upon filling its Roche lobe, that instability will not be manifested until the mass transfer rate exceeds a thermal rate, and becomes too rapid for thermal relaxation to damp its growth. We have therefore adopted an approach to evaluating threshold conditions for dynamical instability that accounts explicitly, if crudely, for the phase of growth in mass transfer rate up to the point at which donor star response can be reasonably approximated as adiabatic.

As a first approximation of the response of the donor star, we suppose that its structure can still be approximated by its adiabatic mass loss sequence, notwithstanding thermal relaxation.  This is the same assumption employed above in illustrating the threshold for a delayed dynamical instability, and is subject to the limitations and shortcomings described there. In a prompt dynamical instability, on the other hand, this should be a much better approximation, as the growth time scale for the mass transfer rate is generally much shorter than a thermal time scale.

The mass loss rate from the donor is a function of the structure of the outer envelope of the donor star, and of the degree to which it overfills its Roche lobe,
\begin{equation}
\dot{M}_1 = \dot{M}_1 (\Delta  R_1)\ .
\label{m1dotdelR}
\end{equation}
Details of the calculation of $\dot{M}_1$ as adopted in this study can be found in the Appendix. At the start of a mass loss sequence, the donor is assumed to fill its Roche lobe exactly:
\begin{equation}
R_{\rm 1i} = R_{\rm L,i}
\label{Rinit}
\end{equation}
As mass transfer proceeds, we assume that the Roche lobe follows the radius-mass relation for conservative mass transfer appropriate to its initial mass ratio, $q_{\rm i} = M_{\rm 1i}/M_{\rm 2i}$.  Defining mass fraction
\begin{equation}
\mu \equiv \frac{q}{1+q}\ ,
\label{massfrac}
\end{equation}
we can write
\begin{equation}
R_{\rm L}(\mu) = R_{\rm 1i} \frac{r_{\rm L}(\mu)}{r_{\rm L}(\mu_{\rm i})} \left(\frac{\mu_{\rm i}}{\mu} \right)^2 \left(\frac{1-\mu_{\rm i}}{1-\mu} \right)^2\ ,
\label{RLmu}
\end{equation}
where clearly $\mu_{\rm i} = q_{\rm i}/(1 + q_{\rm i}) $.  The donor radius $R_1$ follows the adiabatic mass loss sequence, by assumption:
\begin{equation}
R_1(M_1) = R_{\rm ad}(\mu M_{\rm 1i}/\mu_{\rm i})\ ,
\label{R1mu}
\end{equation}
where $M_1 = \mu M_{\rm 1i}/\mu_{\rm i}$ follows from conservation of total mass. The desired threshold condition for dynamical time scale mass transfer then consists in finding that initial mass fraction $\mu_{\rm i}$ (or, equivalently, initial mass ratio $q_{\rm i}$) such that the Roche lobe radius, at its deepest penetration into the envelope of the donor star, suffices to drive mass transfer at just a thermal (Kelvin-Helmholtz) rate. For that rate, we adopt
\begin{equation}
\dot{M}_{\rm KH} = - \frac{GM_{\rm 1i}}{R_{\rm 1i} L_{\rm 1i}}\ ,
\label{MKH}
\end{equation}
where the subscripts ``1i'' refer to the values of the corresponding variables at the beginning of the mass loss sequence.

Let us define $R_{\rm KH}(M_1)$ as the radius within the donor star at which
\begin{equation}
\dot{M}(\Delta_{\rm KH}) = \dot{M}_{\rm KH}\ ,
\label{MDelKH}
\end{equation}
where
\begin{equation}
\Delta_{\rm KH} = R_1(M_1) - R_{\rm KH}(M_1)\ .
\label{DelKH}
\end{equation}
The algorithm we have employed to evaluate Eq.~(\ref{MDelKH}) is detailed in the Appendix to this paper. $R_{\rm KH}$ is then implicitly defined by Eqs.~(\ref{R1mu})-(\ref{DelKH}) as a function of $M_{\rm 1i}$, $\mu$, and $\mu_{\rm i}$.  We likewise define
\begin{equation}
\zeta_{\rm KH} \equiv \left( \frac{{\rm d} \ln R_{\rm {KH}}}{{\rm d} \ln M_1} \right)\ .
\label{zetaKH}
\end{equation}
The our desired solution simultaneously satisfies the equations
\begin{equation}
R_{\rm KH}(M_{\rm 1i},\mu,\mu_{\rm i}) = R_{\rm L}(\mu,\mu_{\rm i})
\label{RKH_RL}
\end{equation}
and
\begin{equation}
\zeta_{\rm KH}(M_{\rm 1i},\mu,\mu_{\rm i}) = \zeta_{\rm L}(\mu)\ .
\label{zetaKH_zetaL}
\end{equation}
$M_{\rm 1i}$ is fixed by the choice of initial model, so these constitute two equations in two unknowns ($\mu$ and $\mu_{\rm i}$).
The solution value for $\mu_{\rm i}$ then corresponds to the minimum initial mass fraction above which the binary suffers dynamical instability; that for $\mu$ specifies the mass fraction at which the instability actually occurs.  The corresponding critical initial mass ratio is $q_{\rm crit} = \mu_{\rm i}/(1 - \mu_{\rm i})$, and we define
\begin{equation}
\zeta_{\rm crit} \equiv \zeta_{\rm L}(q_{\rm crit})
\label{zeta_crit}
\end{equation}
(see Eq.~\ref{zetalobe}) as the initial mass-radius exponent characterizing this critical case. If the binary is subject to a prompt instability, $\mu_{\rm i}$ and $\mu$ differ by a negligible amount, but in a delayed dynamical instability the difference provides an approximate measure of the extent of mass transfer prior to the onset of dynamical instability.

In practice, we have found that this approach works quite satisfactorily, and is insensitive to the vagaries of round-off error that can afflict numerical derivatives like $\zeta_{\rm ad}$ and $\zeta_{\rm KH}$.  However, in the case of donors undergoing pronounced superadiabatic expansion, the ambiguities described in Section~\ref{superadiabatic} lead us to characterize our results somewhat differently.

We conjecture that, when superadiabatic expansion becomes important, the true threshold for dynamical instability probably lies between two extremes.  If we suppose, on the one hand, that a donor only initiates mass transfer when it fills its Roche lobe, as assumed above, then the rapid superadiabatic expansion that follows presents a most favorable circumstance for dynamical instability.  The algorithm detailed above then yields estimates for $q_{\rm crit,min}$ and $\zeta_{\rm crit,min}$, the minimum plausible mass ratio, and corresponding initial mass-radius exponent, unstable to dynamical time scale mass transfer.  If superadiabatic expansion is severe ($\zeta_{\rm KH} \leq -5/3$), no simultaneous solution exists for Eqs.~(\ref{RKH_RL}) and (\ref{zetaKH_zetaL}) in the case of conservative mass transfer.  In this case, $q_{\rm crit,min} \rightarrow 0$, i.e., all mass ratios are dynamically unstable, and we set $\zeta_{\rm crit,min}$ equal to the minimum value of $\ln (R_{\rm KH}/R_{\rm 1i})/\ln (M_1/M_{\rm 1i})$ along the mass loss sequence.

At the opposite extreme, as noted in Section~\ref{superadiabatic}, it is energetically possible for a donor star to initiate mass transfer while its photosphere still lies well within its Roche lobe, given a sufficiently strong perturbation. This circumstance we consider the least prone to dynamical instability, because sustaining mass transfer at a high rate requires rapid contraction of the donor Roche lobe, and therefore a relatively extreme mass ratio. We characterize the initial rapid superadiabatic expansion by $\Delta_{\rm exp}$, the logarithmic difference between the Roche lobe radius and the stellar radius at the onset of mass transfer. To quantify $\Delta_{\rm exp}$, we build a pseudo-envelope model for the donor star in question by assigning to its entire convective envelope a uniform specific entropy equal to that at the base of the convective envelope of the true evolutionary model.  By construction, this pseudo-model has no superadiabatic surface layer.  As seen in Fig.~\ref{GB_expansion} below, the difference between true and pseudo-models is evident only very near the surface.  Then, identifying the radius of this pseudo-model with the Roche lobe radius of the donor, we define
\begin{equation}
\Delta_{\rm exp} = \log (\tilde{R}_{\rm 1i}/R_{\rm 1i})\ ,
\label{Delta_exp}
\end{equation}
where $\tilde{R}_1(M_1)$ is the radius of the pseudo-model as a function of mass. $\Delta_{\rm exp}$ is an estimate of the fraction by which a donor giant branch star may underfill its Roche lobe and yet initiate tidal mass transfer to its companion. In turn, the same algorithm described above for finding $q_{\rm crit}$ and $\zeta_{\rm crit}$, applied now to the pseudo-envelope model, yields estimates for $q_{\rm crit,max}$ and $\zeta_{\rm crit,max}$, the maximum plausible mass ratio, and corresponding initial mass-radius exponent, stable against dynamical time scale mass transfer.

Fig.~\ref{GB_expansion} illustrates the application of the algorithms described above to the determination of threshold conditions for dynamical time scale mass transfer.  In this example, the donor is a $1\ M_{\odot}$ star near the tip of the giant branch.  In response to adiabatic mass loss, the radius of such a donor (heavy solid line) undergoes a rapid initial expansion that will trigger dynamical time scale mass transfer, provided that the mass transfer rate becomes large enough to damp thermal relaxation.  In this example, the donor has a very extended envelope and short thermal time scale, and so it must overfill its Roche lobe by a significant factor in order to drive the mass transfer rate up to $\dot{M}_{\rm KH}$; the depth in the donor at which this occurs is shown by the thin solid line.  The threshold condition for dynamical time scale mass transfer then occurs when the Roche lobe radius just reaches $R_{\rm KH}$. In the example at hand, however, expansion is extremely rapid that all mass ratios are unstable: $\zeta_{\rm crit,min} = -27.4$ and $q_{\rm crit,min} = 0$.  In this critical case, $\dot{M}$ reaches $\dot{M}_{\rm KH}$ at $\log (M_1/M_{\rm 1i}) = -0.00027$, $\log (R_{\rm KH}/R_{\rm 1i}) = 0.00742$; that is, dynamical instability is virtually instantaneous.

Also shown in Fig.~\ref{GB_expansion} is a pseudo-envelope model sequence for the same $1\ M_{\odot}$ giant branch star (heavy dashed line).  The radius in this model at which $\dot{M} = \dot{M}_{\rm KH}$ is shown by the thin dashed line.  In the limiting case, it is energetically possible for this star to sustain Roche lobe overflow by adiabatic expansion when it initially underfills its Roche lobe by $\Delta_{\rm exp}$.  At the threshold for this case  the Roche lobe radius (dotted line) again touches the thin dashed line but, because the donor star underfills its Roche lobe, more dramatic contraction of the Roche lobe is needed to drive $\dot{M}$ up to $\dot{M}_{\rm KH}$.  The corresponding tangent point occurs at $\log (M_1/M_{\rm 1i}) = -0.096$, $\log (\tilde{R}_{\rm KH}/R_{\rm 1i}) = 0.007$, where $\zeta_{\rm L} = 0.455$ and $q = 0.999$, corresponding to $q_{\rm crit,max} = 1.661$ and $\zeta_{\rm crit,max} = 1.928$ at $M_1 = M_{\rm 1i}$.

Our conjecture that the threshold initial mass ratio for dynamical time scale mass transfer, $q_{\rm crit}$, probably lies within the interval $q_{\rm crit,min} \leq q_{\rm crit} \leq q_{\rm crit,max}$, as these limits are estimated above (and likewise for the critical initial mass-radius exponent, $\zeta_{\rm crit,min} \leq \zeta_{\rm crit} \leq \zeta_{\rm crit,max}$) must remain a matter of speculation at present.  It should be noted that the example illustrated in Fig.~\ref{GB_expansion} is a relatively extreme one, and that given the very short thermal time scale of the donor, even dynamically-stable mass transfer on a thermal time scale could trigger common envelope evolution.

\section{Common Envelope Evolution}
\label{CE}

The family of binary stars contains many examples of non-interacting systems containing a compact component (white dwarfs, neutron stars, or black holes) in an orbit so compact that the immediate progenitor of the compact component could not possibly have fit within its Roche lobe. These systems must have originated through common envelope evolution \citep{pacz76}. Examples include planetary nebulae with close binary nuclei, pre-cataclysmic white dwarf-main sequence star pairs, close double white dwarfs, and many binary pulsars. These systems are presumably the progenitors to interacting binaries containing compact components, either through evolution of a non-degenerate stellar component, or through angular momentum loss from the binary via a magnetic stellar wind or gravitational radiation, for example.

Modeling common envelope evolution poses one of the most demanding challenges in computational stellar astrophysics.  The characteristic dimensions of the common envelope may easily exceed those of a compact core embedded within it by factors of $10^4$ or more, corresponding to a ratio of dynamical time scales (envelope versus core) exceeding $10^6$.  Furthermore, the dissipation of energy within the common envelope is governed by ill-understood energy- and angular momentum-transport processes, characterized by a time scale orders of magnitude in excess of the local dynamical time scale, and possibly approaching a stellar thermal time scale.  At present, only the initial immersion of the binary components within a common envelope can be modeled with a semblance of realism (e.g.,~\citet{rick08}).

All is not lost, however, in projecting the outcome of common envelope evolution, as the entire common envelope process must still satisfy conservation laws for energy and angular momentum.  Considering first the energy equation, we can write the initial energy of the binary, $E_{\rm i}$, as
\begin{equation}
E_{\rm i} = - \frac{GM_{\rm 1i}M_{\rm 2i}}{2A_{\rm i}} + E_{\rm 1i} + E_{\rm 2i}\ ,
\label{Ei}
\end{equation}
where $M_{\rm 1i}$ is the initial mass of the donor (with envelope intact), $A_{\rm i}$ is of course the initial binary separation, and $E_{\rm 1i}$ is the initial total energy (gravitational potential plus internal) of the donor star, a negative quantity. Subscripts `2i' refer to the corresponding initial quantities for the companion star.  A similar equation holds for the final, post-common envelope state of the binary (subscripts `f').  As a consequence of common envelope evolution, the envelope of the donor star, mass $M_{\rm env} = M_{\rm 1i} - M_{\rm 1f}$, is shed from the binary.  For reasons outlined in the Introduction, we can assume that the companion star survives common envelope virtually unchanged in mass or total energy,  and so write $M_{\rm 2i} = M_{\rm 2f} = M_2$ and $E_{\rm 2i} = E_{\rm 2f} = E_2$. We require that the final energy, $E_{\rm f}$, of the binary be no greater than its initial energy, $E_{\rm i}$, which yields the inequality
\begin{equation}
\frac{A_{\rm f}}{A_{\rm i}} \leq \frac{M_{\rm 1f}}{M_{\rm 1i}} \left( 1 + \frac{2A_{\rm i} \Delta E_1}{GM_{\rm 1i} M_2} \right)^{-1}\ ,
\label{AfAiE}
\end{equation}
where
\begin{eqnarray}
\Delta E_1 & \!\!\! \equiv \!\!\! & E_{\rm 1f} - E_{\rm 1i} \nonumber \\[4 pt]
& \!\!\! = \!\!\! & \int_{0}^{M_{\rm 1f}} \left(-\frac{Gm}{r} + U\right) {\rm d}m - \int_{0}^{M_{\rm 1i}} \left(-\frac{Gm}{r} + U\right) {\rm d}m
\label{DeltaE1}
\end{eqnarray}
 is the difference between final and initial total energies of the donor star, the initial binding energy of the ejected envelope.  Clearly $M_{\rm 1f} < M_{\rm 1i}$, but it is the fact that the initial binding energy of the envelope usually greatly exceeds the initial orbital energy of the binary in magnitude that is most responsible for dramatic decreases in orbital separation resulting from common envelope evolution.

Considering in turn the conservation of angular momentum, we write the initial angular momentum of the binary, $J_{\rm i}$, as
\begin{equation}
J_{\rm i} = M_{\rm 1i} M_2 \sqrt{\frac{GA_{\rm i} (1 - e_{\rm i}^2)}{M_{\rm 1i} + M_2}}\ ,
\label{Ji}
\end{equation}
where we explicitly allow for initial orbital eccentricity $e_{\rm i} \neq 0$, but neglect the rotational angular momenta of the binary components.  The rotational angular momenta are demonstrably small, typically contributing less than 2\% to the total angular momentum, provided the mass ratio of the binary is not too extreme.  A parallel expression holds for the final angular momentum, $J_{\rm f}$.  Then requiring, as for energy, that the final angular momentum of the binary be no greater  than its initial angular momentum, we find
\begin{equation}
\frac{A_{\rm f}}{A_{\rm i}} \leq \left( \frac{M_{\rm 1i}}{M_{\rm 1f}} \right)^2 \left( \frac{M_{\rm 1f} + M_2}{M_{\rm 1i} + M_2} \right) \left( \frac{1 - e_{\rm f}^2}{1 - e_{\rm i}^2} \right)\ .
\label{AfAiJ}
\end{equation}
In general, dissipative processes prior to, or within, common envelope evolution lead us to expect that $e_{\rm i} \approx 0 \approx e_{\rm f}$.  Clearly, save for the possibility that $\Delta E_1 < 0$, i.e., that spontaneous ejection by the donor star of its envelope is energetically possible, the energy constraint, Eq.~(\ref{AfAiE}), always poses a stronger constraint on post-common envelope orbital separation than does Eq.~(\ref{AfAiJ}).

In the years since common envelope evolution became the generally-accepted model for the origin of compact binaries with an evolved component, various approximations have been put forward for $\Delta E_1$ \citep[e.g.,][]{iben84,webb84,kool90,iben93,webb07}. A more realistic approach \citep[e.g.,][]{han94,han95,dewi00} has been explicitly to integrate over the putative binding energy of the envelope:
\begin{equation}
E_{\rm bind} = \int_{M_{\rm f}}^{M_{\rm i}} \left( - \frac{Gm}{r} + U \right)\,{\rm d}m\ .
\label{Ebind}
\end{equation}
In the limit that this integral extends over the entire mass distribution ($M_{\rm f} \rightarrow 0$), this expression gives the total energy of a star of total mass $M_{\rm i}$; but it is \emph{not} rigorously correct when parsed over a portion of the star ($M_{\rm f} > 0$).  The reason is that the total gravitational potential energy of a star is a global property of that star. The gravitational potential at any point in the star (in the usual convention, where that potential vanishes at infinity) depends not only on the mass interior to that point, but also on the mass exterior to it, which determines the shape (and integrated depth) of the potential outside that point.

It may help to clarify the issue to consider an analytic toy model.  Consider an isentropic $n = 1$ polytrope.  Such a polytrope has the equation of state
\begin{equation}
P = K \rho^2\ ,
\label{eos2}
\end{equation}
where $K$, the polytropic constant, is a function only of entropy, by assumption.  The specific internal energy of a gas obeying this equation of state is
\begin{equation}
u = K \rho\ .
\label{u1}
\end{equation}
An $n=1$ polytrope has the analytic solution \citep[cf.][]{chan39}
\begin{equation}
\theta = \frac{\sin \xi}{\xi}\ ,
\label{theta1}
\end{equation}
where $\xi$ is a dimensionless radial coordinate ($\xi = r/\alpha$), and $\theta$ is the $n$th (i.e. first) root of the ratio of local mass density to central mass density ($\theta = [\rho(\xi)/\rho(0)]^{1/n}$).  The surface of the polytrope corresponds to the first zero of $\theta$, i.e., $\xi_1 = \pi$.  The scale for the radial coordinate is, for an $n=1$ polytrope,
\begin{equation}
\alpha = \left( \frac{K}{2\pi G} \right)^{1/2}\ ,
\label{alpha1}
\end{equation}
and the central density is
\begin{equation}
\lambda = \frac{M}{4\pi^2 \alpha^3}\ .
\label{lambda1}
\end{equation}
Note (cf. Eq.~\ref{alpha1}) that, in the case of an $n = 1$ polytrope, the radius $R = \alpha \xi_1$ is a function only of the polytropic constant $K$, independent of the mass $M$.  The local mass coordinate corresponding to $\xi$ is
\begin{eqnarray}
m &\!\!\!=\!\!\!& - \frac{M}{\pi} \, \xi^2 \frac{{\rm d} \theta}{{\rm d} \xi} \nonumber \\
& = & - \frac{M}{\pi}\, (\xi \cos \xi - \sin \xi)\ .
\label{m1}
\end{eqnarray}

The putative gravitational potential energy per unit mass at radial coordinate $\xi$ is then
\begin{equation}
- \frac{Gm}{r} = \frac{GM}{R} \, \xi \, \frac{{\rm d} \theta}{{\rm d} \xi} \ ,
\label{potl1}
\end{equation}
and the internal energy per unit mass
\begin{equation}
u = \frac{GM}{2R} \, \theta\ ,
\label{Uint1}
\end{equation}
where $M$ and $R$ are the total mass and radius of the polytrope, respectively.  Integrating these energy terms with respect to mass, one finds
\begin{eqnarray}
\Omega(r) &\!\!\!=\!\!\!& \int_0^r - \frac{Gm}{r} \,{\rm d}m = \frac{GM^2}{\pi R} \int_0^\xi \xi'^3 \theta \, \frac{{\rm d} \theta}{{\rm d} \, \xi'} \, {\rm d} \xi' \nonumber \\
& = & \frac{GM^2}{\pi R} \left[ \frac{3}{8} \sin 2\xi - \frac{1}{4} \, \xi \cos 2\xi - \frac{1}{2} \, \xi \right]
\label{Omegaxi}
\end{eqnarray}
and
\begin{eqnarray}
U(r) &\!\!\!=\!\!\!& \int_0^r u\, {\rm d}m = \frac{GM^2}{2\pi R} \int_0^\xi \xi'^2 \theta^2 \, {\rm d} \xi' \nonumber \\
& = & \frac{GM^2}{2\pi R} \left[ \frac{1}{2} \, \xi - \frac{1}{4} \sin 2\xi \right] \ .
\label{Uxi}
\end{eqnarray}
Thus, at the surface of the polytrope ($\xi = \xi_1 = \pi$), we find
\begin{eqnarray}
\Omega(R) &\!\!\!=\!\!\!& - \frac{3GM^2}{4R} \\
U(R) & = & \frac{GM^2}{4R} \ ,
\label{OmegaU}
\end{eqnarray}
giving for the total energy of the polytrope,
\begin{equation}
E_{\rm tot} =  \Omega(R) + U(R) = - \frac{GM^2}{2R} \ .
\label{Etot1}
\end{equation}

Now suppose that all matter outside $\xi = \pi/2$ is removed adiabatically from the polytrope.  This value of $\xi$ corresponds to mass coordinate $m = M/\pi$ and radius coordinate $r = R/2$ in the initial polytrope. Since the specific entropy of the polytrope remains unchanged (by assumption), the polytropic constant $K$, the radial scale factor $\alpha$, and hence the surface radius $R$ all remain unchanged by mass loss.  We then have for the final total energy of the polytrope
\begin{equation}
E_{\rm f} = - \frac{GM^2}{2\pi^2 R} \ ,
\label{Etotf1}
\end{equation}
corresponding to a change in total energy
\begin{equation}
\Delta E = E_{\rm f} - E_{\rm i} = \frac{GM^2}{2R} \left( 1 - \frac{1}{\pi^2} \right) \ .
\label{DeltaE}
\end{equation}
But if we integrate over only the envelope that was removed, we find
\begin{eqnarray}
- \int_m^M \left( - \frac{Gm}{r} + u \right) \, {\rm d}m &\!\!\!=\!\!\!& - \frac{GM^2}{4\pi R} \left[ \sin 2\xi - \xi \cos 2\xi - \xi \right]_{\pi/2}^{\pi} \nonumber \\
& = & \frac{GM^2}{2R} \ ,
\label{Eenv1}
\end{eqnarray}
that is, the total energy of the initial polytrope!  Note that the work done on the polytrope in removing this envelope increment by increment from its surface is
\begin{eqnarray}
W &\!\!\!=\!\!\!& - \int_M^m \frac{Gm'}{R(m')} \, {\rm d}m' = \frac{G}{2R} \left[ M^2 - m^2 \right] \nonumber \\
& = & \frac{GM^2}{2R} \left( 1 - \frac{1}{\pi^2} \right) \ ,
\label{Work1}
\end{eqnarray}
i.e., precisely $\Delta E$, the difference between final and initial total energies of the polytrope.  Indeed, an adiabatic change of mass is only possible through work done on the polytrope.

The difference between Eqs.~(\ref{Eenv1}) and (\ref{DeltaE}) is not particular to this example, but a direct result of hydrostatic readjustment in the mass-losing star.  As mass is stripped away from the surface of the star, a complex readjustment occurs in the interior, as the internal energy of the gas does work against the star's self-gravity to re-establish hydrostatic and virial equilibrium.  Since the surface layers are the coolest part of a star, their loss from the star removes little internal energy directly ---  none at all for a polytropic model, where $u \rightarrow 0$ at the surface. But the loss of matter impacts directly the self-gravitational potential energy of the star.  Re-establishment of virial equilibrium then converts  internal energy from the interior into gravitational potential energy, to power expansion to a new state of hydrostatic equilibrium.

The adiabatic mass loss sequences constructed according to the precepts of this paper permit us, for the first time, to calculate rigorously the change in total energy between initial and final states of the donor star, integrated over their entire interiors.  In common envelope evolution, this is the change in total energy, $\Delta E_1$, that must be supplied by the dissipation of orbital energy and that fixes the largest remnant orbital separation allowed energetically (Eq.~\ref{AfAiE}).

The practical problem of calculating $\Delta E_1$ reliably is somewhat more involved.  Common envelope evolution is of greatest interest, and most prevalent, among binary stars with giant branch donors.  These stars have very compact cores containing a significant fraction of the total stellar mass, and so their cores utterly dominate the evaluation of total energy of these stars.  In evaluating the change in total energy on loss of the extended giant envelope, one is then typically seeking the relatively small difference between large numbers.  We estimate that our typical relative uncertainty in calculating $E_{\rm 1}$ is of order $2 \times 10^{-5}$.  In comparison, the ratio of envelope energy to core energy is (very crudely) of order $M^2/R$ to $M_{\rm core}^2/R_{\rm core}$; with $M_{\rm core} \approx M/2$, and $R_{\rm core} \approx 10^{-4}\,R$, this ratio is $4 \times 10^{-4}$.  Numerical limitations to the brute force calculation of the total stellar energy therefore severely limit the utility of that approach in just the circumstances when common envelope evolution is of greatest interest.

Fortunately, the observation above that one can also calculate the change in energy of the donor by tracking the energy content of matter removed from its surface provides a suitable alternative approach to this problem.  Therefore, in addition to calculating $\Delta E_1$ by direct integration of the total energy of initial and final states (Eq.~\ref{DeltaE1}), we can calculate the change in energy of the donor star by integration along the mass loss sequence:
\begin{equation}
\Delta E_1' = \int_{M_{\rm 1i}}^{M_{\rm 1f}} \left( - \frac{GM}{R} + U(R) \right)\, {\rm d}M\ ,
\label{DeltaE1a}
\end{equation}
where the integrand is evaluated at the surface ($r = R$, $m = M$) of the model.  This expression differs from Eq.~(\ref{Work1}) only in that we must account for the finite internal energy of matter at the surface of the model; that contribution vanishes at the surface of a classical polytrope.  Provided we remove small enough mass increments along an adiabatic mass loss sequence to track the change in radius smoothly, we achieve a typical relative uncertainty in $\Delta E_1'$ of order $7 \times 10^{-4}$.   Thus, Eq.~(\ref{DeltaE1a}) affords the more accurate measure of the change of energy of the donor star early in a mass loss sequence, when $|\Delta E_1/E_{\rm 1i}| \lesssim 1/40$, while Eq.~(\ref{DeltaE1}) gives a more accurate measure later in the sequence.  In practice, we find the two estimates are in close agreement at the transition, and use a weighted mean of the two to determine best estimates of $\Delta E_1$ and $E_{\rm 1f}$.

Given an initial donor of mass $M_{\rm 1i}$ and radius $R_{\rm 1i}$, knowledge of $\Delta E_1(M_{\rm 1f})$ allows us to constrain the range of companion masses, $M_2$, and post-common envelope orbital separations, $A_{\rm f}$, that are energetically allowed.  We assume that the initial donor fills its Roche lobe:
\begin{equation}
A_{\rm i} = \frac{R_{\rm 1i}}{r_{\rm L}(M_{\rm 1i}/M_2)} \ .
\label{Ai}
\end{equation}
Then the energy constraint (Eq.~\ref{AfAiE}) gives the largest post-common envelope orbital separation allowed by conservation of energy, for a post-common envelope (remnant) donor of mass $M_{\rm 1f}$:
\begin{equation}
A_{\rm f} \leq A_{\rm i} \frac{M_{\rm 1f}}{M_{\rm 1i}} \left( 1 + \frac{2A_{\rm i} \Delta E_1}{GM_{\rm 1i} M_2} \right)^{-1} \ .
\label{AfE}
\end{equation}
To terminate common envelope evolution, we require that the remnant donor fit within its Roche lobe,
\begin{equation}
A_{\rm f} \geq \frac{R_{\rm 1f}}{r_{\rm L}(M_{\rm 1f}/M_2)} \ ,
\label{Af1}
\end{equation}
and that the companion star also fit within its Roche lobe,
\begin{equation}
A_{\rm f} \geq \frac{R_2}{r_{\rm L}(M_2/M_{\rm 1f})}\ .
\label{Af2}
\end{equation}

\begin{figure}
\begin{center}
\resizebox{16cm}{!}{\rotatebox{-90}{\includegraphics{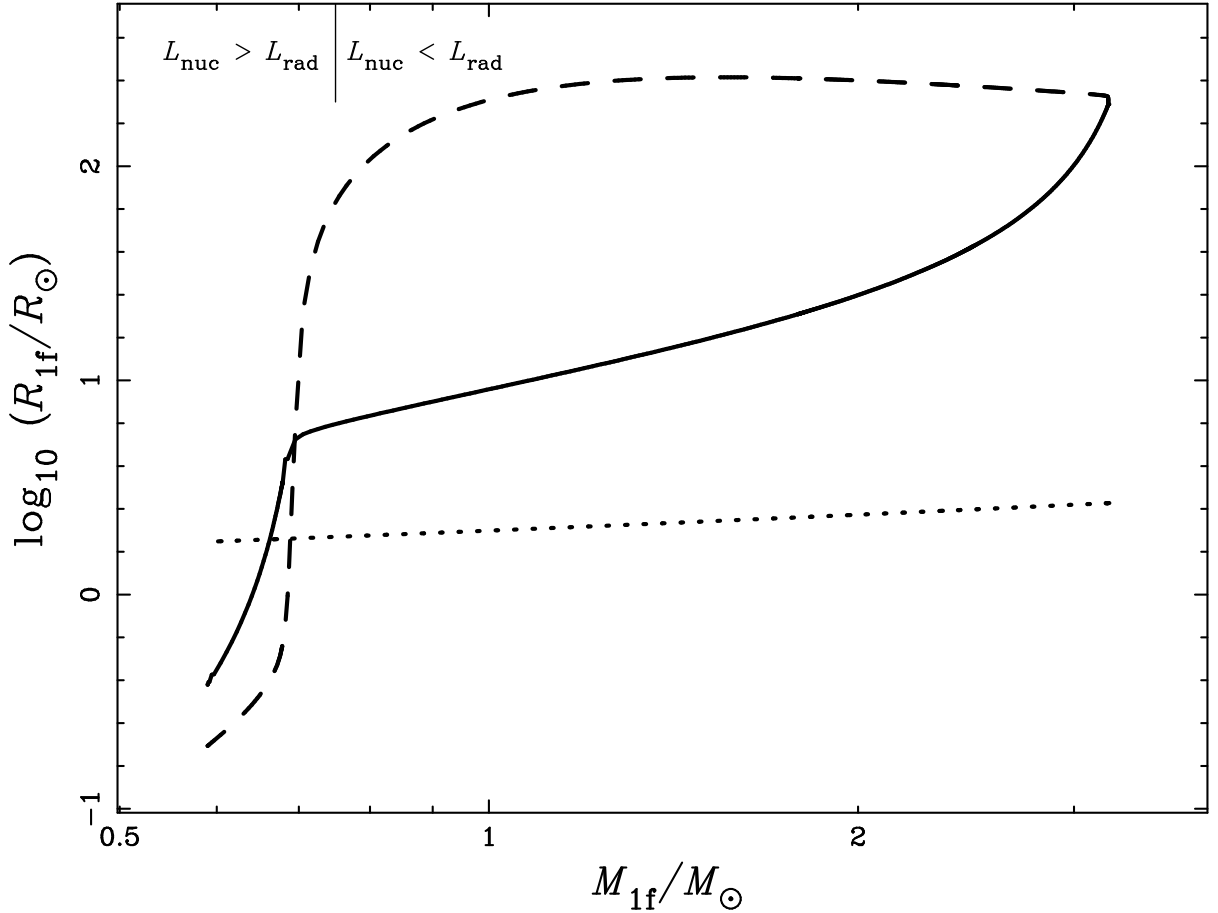}}}
\caption{Limiting remnants of common envelope evolution.  In this example, the donor is a $3.2\ M_{\odot}$ asymptotic-branch donor star, and its companion a $0.8\ M_{\odot}$ main sequence star.  The solid line marks the maximum post-common-envelope separations allowed by energy conservation (Eq. \ref{AfE}), the dashed line the minimum separation needed to accommodate the donor within its Roche lobe (Eq.~\ref{Af1}, and the dotted line the minimum separation needed to accommodate the companion star.  No systems with $M_{\rm f} > 0.690\ M_{\odot}$ survive, because the donor star radius is still too large to fit within its Roche lobe, while systems with $M_{\rm f} < 0.663\ M_{\odot}$ are too compact to accommodate the companion star.  Note also that remnant donors with with masses $M_{\rm 1f} < 0.75\ M_{\odot}$ have nuclear luminosities exceeding their radiated luminosities; they must expand beyond the dashed line to reach thermal equilibrium, and so may drive a further phase of thermal time scale mass transfer.}
\label{3_2msun_gb_A}
\end{center}
\end{figure}

Fig.~\ref{3_2msun_gb_A} illustrates an application of these constraints on survival of a binary with a 3.2 $M_{\odot}$ asymptotic giant branch donor with a 0.80 $M_{\odot}$ companion. For that assumed companion mass, this figure shows as a function remnant donor mass, $M_{\rm 1f}$ the binary separation from Eq.~(\ref{AfE}), the binary separation required to accommodate the donor star within its Roche lobe from Eq.~(\ref{Af1}), and the corresponding separation required to accommodate the companion star within its Roche lobe from Eq.~(\ref{Af2}).  The binary separation must exceed the limits posed by the Roche lobe radii for survival of common envelope evolution.  Survival of that phase does not necessary signal the end of mass transfer, however, only that any further mass transfer occur on a thermal time scale or longer.  We can ascertain whether the donor likely drives continued mass transfer on its thermal time scale upon emergence from the common envelope by comparing its nuclear luminosity with its surface luminosity.  The difference between nuclear and radiated luminosities must be made up work done by expansion or contraction within the star; if the nuclear luminosity exceeds the surface luminosity (see Fig.~\ref{3_2msun_gb_A}), net energy absorption in the stellar interior is expected to drive further thermal time scale expansion of the donor.

\section{Conclusions and Outlook}
\label{conclusions}

In this paper we have described the construction and interpretation of model sequences describing the asymptotic responses of stars to mass loss in a binary system in the adiabatic limit, in which mass transfer is so rapid that negligible heat transport occurs within the stellar interior during mass loss.  We have shown how the radial responses revealed by these adiabatic mass loss sequences can be used to establish threshold conditions for dynamical time scale mass transfer.  That instability may occur promptly upon the donor filling its Roche lobe, if the donor has a surface convection zone of any significant depth, or following an extended period of thermal time scale mass transfer (a delayed dynamical instability) when a radiative donor is stripped down to a nearly isentropic core.  Among very luminous stars with deep surface convection zones (bright giant branch and asymptotic giant branch stars), the adiabatic mass loss sequences show a pronounced initial expansion, as their superadiabatic surface layers are stripped away, complicating the interpretation of these models.

Adiabatic mass loss sequences can also be used to constrain the survival of binaries entering common envelope evolution through dynamical time scale mass transfer.  We correct certain conceptual errors appearing in the literature regarding the calculation of envelope binding energy, and describe complementary approaches to its evaluation.  By combining energy constraints with requirements that both binary components fit within their post-common envelope Roche lobes, we can place strict limits on the masses, mass ratios, and remnant orbital separations of binaries passing through common envelope evolution.

In future papers we will survey the Hertzsprung-Russell diagram, mapping out the the threshold conditions for dynamical time scale mass transfer, and for the survival of common envelope evolution, across the full range of masses and evolutionary stages of potential interest, from main sequence stars across the Hertzsprung gap, to giant branch stars and advanced stages of nuclear burning.  These are essential considerations for binary population synthesis models \citep[e.g., ][]{han94,han95,port96,lip96,hurl02,tut02,will05,belcz08}

\acknowledgments

We are indebted to the referee, Chris Tout, for a very careful, thorough, and constructive review of this paper. His effort has been invaluable to us in seeking greater clarity. This work was supported by grants from the National Natural Science Foundation of China (10821061, 10973036 and 2007CB815406), the U.S. National Science Foundation (AST 0406726), the Chinese Academy of Sciences (KJCX2-YW-T24), and the Yunnan Natural Science Foundation (08YJ041001).

\appendix{\noindent \bf Appendix: Roche Lobe Overflow}
\renewcommand{\theequation}{A\arabic{equation}}

We calculate the mass transfer rate from a lobe-filling star by integrating over a plane perpendicular to the line of centers connecting the two binary components, and passing through the inner Lagrangian ($L_1$) point. The mass transfer stream passes through a sonic surface which must coincide closely with this surface of integration \citep[cf.][]{lub75,egg06}. Thus
\begin{eqnarray}
\dot{M}_1 & \!\!\!=\!\!\! & - \int_{L_1} \rho \mathbf{v} \cdot \, {\rm d} \mathbf{\Sigma} = - \int_{\rm L_1} \rho c_s \, {\rm d} \Sigma \nonumber \\
& \!\!\!=\!\!\! & - \int_{\rm L_1} \rho c_s \frac{{\rm d} \Sigma}{{\rm d} \phi} \, {\rm d} \phi \ ,
\label{mflux}
\end{eqnarray}
where $\rho$ and $c_s$ are the local gas density and sound speed, respectively, and the integral is taken over the sonic surface (area element ${\rm d} \mathbf{\Sigma}$), which we assume to coincide with the plane of integration defined above. Taylor expansion of the Roche potential about the inner Lagrangian point, $L_1$, point gives, to lowest order~\citep{webb77},
\begin{equation}
\phi - \phi_{\rm L} = \frac{G(M_1+M_2)}{A} \left[ - \frac{1}{2} (2a_2 + 1) (x - x_{\rm L})^2
+ \frac{1}{2} (a_2 - 1) y^2 + \frac{1}{2} a_2 z^2 + \cdots \right] \ ,
\label{phi_Roche}
\end{equation}
where $A$ is the binary separation; ($x$,$y$,$z$) form a right-handed coordinate system (in units of $A$), centered on the primary (star 1), with the $x$-axis directed toward the secondary (star 2), $y$-axis in the direction of orbital motion of star 1, and $z$-axis parallel to the orbital angular momentum vector; ($x_{\rm L},0,0$) are the coordinates in this system of $L_1$, and $a_2$ is
\begin{equation}
a_2 = \frac{\mu}{x_L^3} + \frac{(1 - \mu)}{(1 - x_L)^3} \ ,
\label{a2}
\end{equation}
where $\mu \equiv M_1/(M_1+M_2) = q/(1+q)$, as before.  A simple, accurate approximation to $x_{\rm L}$ is
\begin{equation}
x_{\rm L} =  \left\{
\begin{array}{ll}\left(0.7-0.2q^{1/3}\right)q^{1/3} & \mbox{if}\ q \le 1, \\
1-\left(0.7-0.2q^{-1/3}\right)q^{-1/3}   & \mbox{if}\ q \ge 1.
\end{array} \right.
 \label{xL}
\end{equation}
Surfaces of constant $\phi$ are, to lowest order, concentric ellpsoids. The intersection of these with the plane $x = x_{\rm L}$ are concentric ellipses with axial ratio $(a_2 - 1)/a_2$, and cross-sectional area $\Sigma$ increasing linearly with potential $\phi$ so that
\begin{equation}
\frac{{\rm d} \Sigma}{{\rm d} \phi} = \frac{2\pi A^3}{G(M_1 + M_2)[a_2(a_2 - 1)]^{1/2}} \ .
\label{dSdphi}
\end{equation}

We assume for simplicity that the mass outflow from the donor star is laminar, and occurs along equipotential surfaces.  Let $P$ and $\rho$ be the pressure and density on a given equipotential far from $L_1$, where the material is almost stationary.  We assume that they follow a power-law adiabat,
\begin{equation}
P = K \rho^{\Gamma_1}\ ,
\label{adiabat}
\end{equation}
where $K$ and $\Gamma_1  \equiv (\partial \ln P/\partial \ln \rho)_s$ are constant along streamlines.  Then at the sonic surface ($x = x_{\rm L}$), the density $\rho_{\rm L}$ and sound speed $c_{\rm L}$ on the streamline are \citep[cf.][]{webb77}
\begin{equation}
\rho_L = \left( \frac{2}{\Gamma_1 + 1} \right)^{1/(\Gamma_1 - 1)} \rho\ .
\label{rhoL}
\end{equation}
and
\begin{equation}
c_{\rm L} = \left( \frac{2\Gamma_1}{\Gamma_1 + 1}\,K \rho^{\Gamma_1 - 1} \right)^{1/2} = \left( \frac{2\Gamma_1}{\Gamma_1 + 1}\, \frac{P}{\rho}  \right)^{1/2}\ .
\label{cL}
\end{equation}
Combining Eqs.~(\ref{mflux}), (\ref{dSdphi}), (\ref{rhoL}), and (\ref{cL}), we  can write
\begin{equation}
\dot{M}_1 = - \frac{2 \pi R_L^3}{G M_1} f(q)  \int_{\phi_{\rm L}}^{\phi_s} \Gamma_1^{\frac{1}{2}} \left( \frac{2}{\Gamma_1 + 1} \right)^{\frac{\Gamma_1 + 1}{2(\Gamma_1 - 1)}}\,  ( \rho P)^{\frac{1}{2}}\, {\rm d} \phi \ ,
\label{m1dot}
\end{equation}
where $\phi_s$ is the stellar surface potential, and we have used the relation $R_{\rm L} = r_{\rm L} A$. We have combined factors dependent on the mass ratio in the coefficient
\begin{equation}
f(q) \equiv \frac{q}{r_{\rm L}^3 (1+q)}\frac{1}{[a_2 (a_2 - 1)]^{1/2}}\ ,
\label{fq}
\end{equation}
a slowly-varying function of $q$.  For the integration over potential $\phi$, we use the approximation
\begin{equation}
{\rm d}\phi = \frac{GM_1}{R^2}\,{\rm d}R\ ,
\label{dphi}
\end{equation}
and integrate from $R_{\rm L}$ to the stellar radius $R$.

Eq.~(\ref{m1dot}) is at best a rough approximation.  In addition to approximation of the equation of state by a simple power law, neither of the assumption of laminar flow nor that of flow along equipotential surfaces is valid.  A substantial fraction of the mass outflow wells up from within the Roche lobe of the donor, rather than flowing along its surface (see, e.g.,~\citet{egg06}).  However, we find that the adiabatic index $\Gamma_1$ often varies dramatically with depth through the outer envelope of the donor star, and when $\Gamma_1 \leq 1$ (as often happens in partial ionization zones) no solution for an irrotational outflow exists. The outflow must then be turbulent.  The assumptions of laminar flow and equipotential flow err in opposite senses. A significant fraction of the work done by pressure gradients is spent in driving turbulence rather than bulk flow, reducing the mass transfer rate, while the upwelling of gas from the interior increases it.  We consider that our estimate for $\dot{M}_1$ above is accurate within a factor of two or so --- for constant $\Gamma_1 > 1$, it yields the same functional dependence on $M_1$, $R_{\rm L}$, and $\phi_{\rm s} - \phi_{\rm L}$ as does \citeauthor{egg06}'s \citeyearpar{egg06} prescription --- but a rigorous treatment will require full three-dimensional models of the the outflow.

\begin{figure}
\begin{center}
\resizebox{16cm}{!}{\rotatebox{-90}{\includegraphics{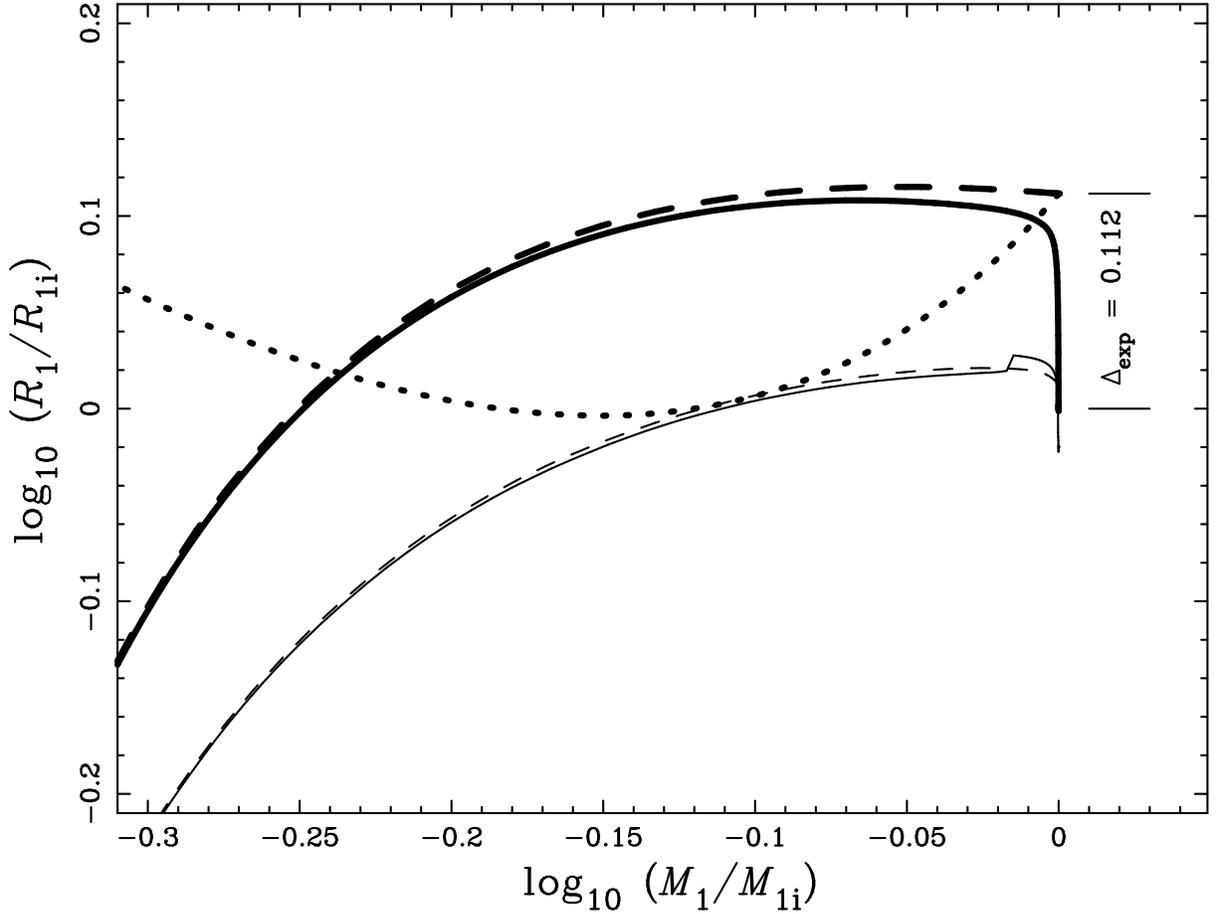}}}
\caption{Adiabatic mass loss sequences for true (thick solid lines) and pseudo-models (thick dashed lines) of a $1\ M_{\odot}$ star near the tip of the giant branch. The pseudo-models have isentropic envelopes, enabling us to quantify $\Delta_{\rm exp}$, the prompt initial expansion resulting from removal of superadiabatic surface layers. Also shown are the Roche lobe radii within true and pseudo-models (thin solid and dashed lines, respectively) at which the mass loss rate would reach $\dot{M}_{\rm KH}$. At higher mass loss rates, thermal relaxation freezes out.  Critical initial mass ratio at which $\dot{M}_1$ just reaches $\dot{M}_{\rm KH}$ for the pseudo-model is shown by the heavy dotted. (The corresponding limit for the true model is omitted for clarity). We believe that these two measures of the critical mass ratio probably bracket the true threshold for dynamical instability.}
\label{GB_expansion}
\end{center}
\end{figure}

\end{document}